\documentclass[]{elsart}
\usepackage{amsmath}
\usepackage[]{graphicx}
\usepackage{dsfont}
\graphicspath{{figures/}}
\newcommand{\MSbar}{\overline{\text{MS}}}
\newcommand{\MStilde}{\widetilde{\text{MS}}}
\DeclareMathOperator{\FT}{FT}

\newcommand{\als}{\alpha_s}
\newcommand{\as}{a_s}

\newcommand{\re}[1]{(\ref{#1})}

\newcommand{\beq}{\begin{equation}}
\newcommand{\eeq}{\end{equation}}
\newcommand{\bea}{\begin{eqnarray}}
\newcommand{\eea}{\end{eqnarray}}

\newcommand{\ice}[1]{\relax}
\newcommand{\dmu}{\mu^2\frac{\text{d}}{\text{d}\mu^2}}

\begin{document}
\begin{frontmatter}
\begin{flushleft}
TTP10-42\\
SFB/CPP-10-89
\end{flushleft}
\title{Massless correlators of vector, scalar and tensor currents
in  position space at   orders $\alpha_s^3 $  and $\alpha_s^4$ :
explicit analytical results
}
\author[Karlsruhe,Moscow]{K. G. Chetyrkin},
\author[Karlsruhe]{A. Maier},
\address[Karlsruhe]{Institut f\"ur Theoretische Teilchenphysik,
Karlsruhe~Institute~of~Technology~(KIT), 76128 Karlsruhe, Germany}
\address[Moscow]{Institute for Nuclear Research, Russian Academy of
  Sciences, Moscow 117312, Russia}
\begin{abstract}
We present  analytical results both in  momentum and
position  space for the massless correlators of the vector and
scalar currents to order $\alpha_s^4$ as well as for the tensor
currents to order $\alpha_s^3$. The evolution
equations for the correlators together with   all relevant
anomalous dimensions are discussed in  detail.
As an  application we present explicit conversion formulas  relating
the $\MSbar$-renormalized vector, scalar and tensor currents to their counterparts
renormalized in the X-space renormalization scheme more appropriate for lattice
calculations.
\end{abstract}

\end{frontmatter}

\section{Introduction}
\label{sec:intro}

Correlators of gauge invariant quark currents are important objects in QCD. 
It is  enough to mention that the correlator of  two vector currents
is directly related to the famous ratio
$
R(s) = {\sigma(e^+e^-\to {\rm hadrons})\over \sigma(e^+e^-\to
    \mu^+\mu^-)}\,
$,  the Adler function 
and the  decay widths  of the Z-boson and the $\tau$-lepton (for a review see, e.g.
\cite{Chetyrkin:1996ia}). 

To be specific, let us consider a correlator: 
\begin{equation}
  \label{generic}
  \Pi(q)=i \int \text{d}x\ e^{iqx}\langle T j(x)j^{\dagger}(0)\rangle
{},
\end{equation}
with $j$ being a  gauge invariant local operator.
The polarization operator $\Pi(q)$ satisfies the standard dispersion 
relation\footnote{For simplicity we assume that 
the current  $j$ is a Lorentz scalar.}
\begin{equation}  \label{disp}
  \Pi(q)= \int_0^\infty  \text{d}s  \frac{\rho(s)}{s-q^2}
-\mbox{subtractions}
{},
\end{equation}
The subtractions on the right-hand side of Eq.~\re{disp} are necessary as they
remove an additional divergence coming from the vicinity of the region
$x \sim 0$ in the $x$-integration in \re{generic}. The structure of
the correlator \re{generic} is significantly simplified
if the momentum $q$ is considered as large compared to (active) quark masses.
Setting then all quark masses to zero one  can describe the general structure
of the correlator in pQCD as follows:
\[
\Pi(q) = (Q^2)^{d -2}\sum_{i=1}^{\infty} \sum_{i \ge k \ge 0} \Pi_{ik}\, a_s^{i-1}(\mu) 
\,\left(\ln\frac{\mu^2}{Q^2}\right)^k
{}.
\label{gen:structure}
\]
Here $Q^2 \equiv -q^2$, $a_s = \frac{\als}{\pi}$, $\mu$ stands for the
renormalization scale and $d$ is the  (mass) dimension of the
current $j$.

In some applications it is useful to deal with the correlators in position 
space (see, e.g. 
\cite{Chetyrkin:1988yr,Shuryak:1993kg,Schafer:1996wv,Schafer:2000rv,Narison:2001ix} 
and below).  Using  a text-book 
formula for the massive scalar propagator in position space: 
\beq
\Delta(x,s) \equiv \frac{1}{i (2\pi)^4} \int_0^\infty \text{d} q   
 \,  \frac{e^{iqx}}{ s - q^2} = \frac{1}{-4\,\pi^2 x^2} \, z\, K_1(z), \ \ z = \sqrt{-x^2\, s},
\eeq
(with $K_1$ being a modified Bessel function)
we arrive at a well-known representation for $\Pi$:
\beq
\Pi(x) =  \int_0^\infty  \text{d}s\, \Delta(x,s)\, \rho(s)
{}.
\eeq
It should be stressed that the spectral density $\rho(s)$ does not depend
on the non-logarithmical contributions to the sum in \re{gen:structure} (that is
those proportional to the coefficients $\Pi_{ik}$ with $ k\equiv 0$).
Thus, the full correlator in position space considered as a 
function of $x$ (defined for all $x$ with $x^2 \not = 0$ ) also does not
depend on non-logarithmical contributions to $\Pi(q)$. 

In general, the  operator $j$ is not scale-invariant (equivalently, 
has a non-zero anomalous dimension). The renormalization of the    operator
and the position space correlator look as follows\footnote{Note that the momentum space correlator 
is renormalized in a more complicated way due to the UV divergence at small $x$. The corresponding
formula is given below in  Section \ref{sec:def}.}
\beq
j = Z_j \, j_0, \ \ \  \Pi(x) = Z^2_j \,\Pi_0(x)
{},
\eeq
where $j_0$ and $\Pi_0$ stand for the corresponding bare quantities.

An important feature of the position space correlators is that 
they can be directly computed non-perturbatively on the    lattice 
by Monte Carlo  simulations (see, e.g. 
\cite{Chu:1993cn,DeGrand:2001tm,Gimenez:2004me,Gimenez:2005nt}). Their long-distance
behavior is  governed  by the non-perturbative features of the  underlying 
field theory, QCD. On the other  hand, due to asymptotic freedom,
their short-distance behavior can be described by perturbation theory and 
operator product expansion (OPE).
A  meaningful comparison of perturbative results at short distances with 
their lattice counterparts requires, obviously,  the use of 
{\em one and  the same}  renormalization prescription in the common case of scale-dependent  
operators. While {\em minimal subtraction} schemes ($\MSbar$ 
and its relatives \cite{'tHooft:1973mm,Bardeen:1978yd,Chetyrkin:1980pr}) are certainly
 preferable for perturbative calculations, they, clearly, can not be implemented  on lattice.

A solution of the problem 
is based on the use of an intermediate renormalization scheme, with the
renormalization conditions imposed directly on quark and gluon Green
functions computed in a fixed gauge and for a particular configuration
of external momenta \cite{Martinelli:1994ty}.

A convenient intermediate scheme for the renormalization of the quark
current operators has been developed in \cite{Gimenez:2004me}. It is
based on the study of the corresponding position space correlators and
is called the X-space scheme. The conversion formulas between the
$\MSbar$ and the X-space scheme have been elaborated in
\cite{Gimenez:2004me} to the next-to-leading order.

Recently there has been a lot of progress in computing higher order
corrections to the vector  and scalar  correlators within
perturbative QCD both in the massless limit as well as for the general
case of massive quarks.  Both correlators are now known in momentum
space to order $\als^3$ \cite{GorKatLar91:R(s):4l,SurSam91,Chetyrkin:1996sr,Kiyo:2009gb} 
(the real and absorptive parts) and
even, partially, to order $\als^4$ \cite{Baikov:2005rw,Baikov:2008jh} (only the absorptive part
in the massless limit). The situation is not so good for the tensor
correlator, which is known completely to order $\als^2$ in the
massless limit only \cite{Gracey:2009da}.

In addition, the anomalous dimensions of the scalar and tensor
currents are known to order $\als^4$
\cite{Chetyrkin:1997dh,Vermaseren:1997fq,Baikov:2006ai} (vector and
axial-vector currents have identically vanishing anomalous dimension due to the
corresponding Ward identity).


The aims of the present paper are:
\begin{itemize}

\item To compute the order  $\als^3$ contribution to the tensor correlator 
(only  absorptive  part in the massless limit).

\item To summarize available momentum space results for the  scalar, 
vector and tensor (all massless) correlators and corresponding
anomalous dimensions.

\item To discuss in  detail the evolution equations for  all three correlators.

\item To present {\em  full} results for the  correlators in position space, namely: scalar and  vector to (and including)  order 
$\als^4$ and tensor to (and including) order $\als^3$ computed within 
massless QCD.

\item To construct the $N^4LO$ conversion formulas between $\MSbar$ and X-space 
renormalized scalar and  vector currents as well as the $N^3LO$ ones for the tensor current.

\item To study the stability of the conversion formulas with respect 
to higher order (not yet computed) perturbative corrections. 
      
\end{itemize}

The plan of the paper is as follows. In the next section we discuss
our conventions and the definition of the X-space renormalization
scheme as well as a version of the $\MSbar$-scheme --- the
$\widetilde{\mbox{MS}}$-one which seems to be more convenient for
renormalization of the position space correlators. Sections \ref{sec:q_space_res} and \ref{sec:x_space_res}
list all available results for (massless) quark currents correlators
in momentum and position space respectively. In Section \ref{comments}
we try to provide the reader with the concise bibliographical
information about the origin of the results collected in the two previous Sections
as well as about the main technical tools employed in
the corresponding calculations. Conversion formulas between the X-scheme
and $\MStilde$/$\MSbar$ schemes are discussed in Section \ref{sec:conv}.  In
the last section \ref{sec:conc} we summarize the content of the
paper. 

In addition, there are three appendixes. 
In Appendix~\ref{sec:min_to_eucl} we spell out the rules which we use to construct
the Euclidean correlator from its  Minkowskian counterpart.
Appendix~\ref{sec:FT} provides the reader
with necessary information on the Fourier transformation. Appendix~\ref{sec:anom_dim} lists
various  anomalous dimensions relevant   for the RG evolution  
of the quark current correlators.

\section{Quark current correlators in momentum and position space}
\label{sec:def}

In this section we outline  our conventions and recall the definition of
the X-space scheme as presented in
Ref. \cite{Gimenez:2004me}. Our discussion will focus on the
correlator of scalar currents first. The generalization to other
Lorentz structures is straightforward, we will comment on it towards
the end of the section.

The scalar correlator in momentum space is defined as 
\begin{equation}
  \label{eq:s_corr_q}
  \Pi^S(q)=i \int \text{d}x\ e^{iqx}\langle Tj(x)j(0)\rangle
\end{equation}
with $j=\bar{\psi}\mathds{1}\psi$. For space-like momenta we can express
the correlator in terms of the Euclidean momentum $Q$. In what  follows we will work exclusively with Euclidean correlators. Our 
procedure of obtaining 
Euclidean  correlators  from  Minkowskian ones is described in Appendix~\ref{sec:min_to_eucl}. 
The correlator considered in
position space reads
\begin{equation}
  \label{eq:s_corr_x} 
  \Pi^S(X)=\langle j(X)j(0)\rangle
\end{equation}
with  a Euclidean separation $X$.
We work in the chiral limit with $m_\psi =0$. Note that diagrams with
purely gluonic cuts do not contribute to the scalar correlator (in the assumed massless limit) .

\subsection{Momentum space}
\label{sec:def_q_space}

We denote the $\MSbar$ renormalized momentum space correlator at the
scale $\mu$ by
\begin{equation}
  \label{eq:def_s_MS}
  \Pi^S(Q,\mu)=(Z^S)^2\Pi_0^S(Q)+Z^{SS}(\mu^2)^{-\epsilon}Q^2\,,
\end{equation}
where $\Pi_0^S(Q)$ is the bare scalar correlator.
Note that in addition to the multiplicative renormalization with $Z^S$
there is a subtractive counterterm $Z^{SS}$. The corresponding
renormalization group equation reads
\begin{equation}
  \label{eq:RGE}
  \mu^2\frac{\text{d}}{\text{d}\mu^2}\Pi^S(Q,\mu)=2\gamma^S\Pi^S(Q,\mu)+\gamma^{SS}Q^2
\end{equation}
with the anomalous dimensions
\begin{equation}
  \label{eq:def_gamma}
  \gamma^S=\mu^2\frac{\text{d}\log Z^S}{\text{d}\mu^2}\,,\qquad
  \gamma^{SS}=\mu^2\frac{\text{d} Z^{SS}}{\text{d}\mu^2}-(2\gamma^S+\epsilon) Z^{SS}\,.
\end{equation}
Using the solution to the renormalization group equation (\ref{eq:RGE}), we can evolve
the correlator from one scale $\mu_0$ to a different scale
$\mu_1$:
\begin{align}
  \label{eq:RGE_q}
  \Pi^S(Q,\mu_1)=&
\exp\left(\;\int\limits_{a_s(\mu_0)}^{a_s(\mu_1)}\frac{\text{d}z}{z}
\frac{2\gamma^S(z)}{\beta(z)}\right)\big(\Pi^S(Q,\mu_0)+Q^2\Delta(\mu_1,\mu_0)\big)
\,,\notag\\
\Delta(\mu_1,\mu_0)=&\int\limits_{a_s(\mu_0)}^{a_s(\mu_1)}
\frac{\text{d}z}{z}\frac{\gamma^{SS}(z)}{\beta(z)}
\exp\left(-\int\limits^z_{\as(\mu0)}\frac{\text{d}z'}{z'}\frac{2\gamma^S(z')}{\beta(z')}\right)\,,
\end{align}
where $a_s = \alpha_s/\pi= g^2/(4\pi^2)$, $g$ is the  strong coupling constant and 
the $\beta$-function $\beta(a_s)$ is defined as 
\beq
\dmu a_s  = \as \beta(a_s) \equiv
-\sum_{i\geq0}\beta_i a_s^{i+2}
\label{beta}
{}.
\eeq
\ice{
\begin{equation}
  \label{eq:a_s}
  a_s(\mu)=\frac{\alpha_s(\mu)}{\pi}\,.
\end{equation}
}
While it is of course possible to recover logarithms explicitly
using this solution, it is more convenient to rewrite
the renormalization group equation into a differential equation in
$l_{\mu Q}=\log(\mu^2/Q^2)$ for this purpose:
\begin{equation}
  \frac{\partial}{\partial l_{\mu Q}}\Pi^S(Q)=2\gamma^S\Pi^S(Q)+\gamma^{SS}Q^2-\beta\,a_s\frac{\partial}{\partial a_s}\Pi^S(Q)\,,
  \label{eq:rge_log}
\end{equation}
This equation can be used to iteratively reconstruct the logarithmic parts of $\Pi^S(Q)$.
Explicit formulas for the anomalous dimensions and the QCD $\beta$
function are given in Appendix \ref{sec:anom_dim}.

\subsection{Position space}
\label{sec:def_x_space}

In principle, the discussion of the preceding paragraph can be directly
translated to the position space correlator. It is, however, convenient
to use a modification of the MS scheme that is a bit different from the
traditional $\MSbar$ convention. The reason for this  is that in
 the $\MSbar$ scheme logarithms in position space naturally appear in the
 form\footnote{See Appendix \ref{sec:FT_log} for more details.}
 \begin{equation}
   \label{eq:log_MS_x}
   \log\left(\frac{\mu^2X^2}{4}\right)+2\*\gamma_E\,.
 \end{equation}
We can transform these to the simpler form $\log(\mu^2X^2)$ with a shift in the renormalization scale:
\begin{equation}
  \label{eq:mu_to_mu}
  \mu \to 2e^{-\gamma_E} \mu \approx 1.12\, \mu\,.
\end{equation}
The shifted $\mu$ defines a new modified MS scheme which we call
$\MStilde$. The relation between $\MStilde$ quantities and their $\MSbar$
counterparts is of course very simple:
\begin{equation}
  \label{eq:def_MStilde}
  \widetilde{\Pi}^S(X,\mu)=\Pi^S\Big(X,2e^{-\gamma_E}\mu\Big)\,,\qquad
  \widetilde{a}_s(\mu)=a_s\Big(2e^{-\gamma_E}\mu\Big)\,,
\end{equation}
Using the evolution of the strong coupling constant we can also relate
the $\MStilde$ coupling to the $\MSbar$ coupling at the same scale:
\begin{equation}
  \label{eq:astilde}
  \begin{split}
    \tilde{a}_s(\mu)=&a_s(\mu)\bigg\{1-a_s(\mu)l\beta_0
      +a_s^2(\mu)l(\beta_0^2l-\beta_1)
      +a_s^3(\mu)l\left(-\beta_0^3l^2+\frac{5}{2}\beta_0\beta_1l-\beta_2
      \right)
    \\
    &+
      a_s^4(\mu)l\left[\beta_0^4l^3-\frac{13}{3}\beta_0^2\beta_1l^2
        +3\left(\frac{\beta_1^2}{2}+\beta_0\beta_2\right)l-\beta_3\right]
      +{\cal O}(a_s^5l^5)\bigg\}\,,
  \end{split}
\end{equation}
where $l=2\,(\log(2)-\gamma_E)$.

In position space there is no additional subtractive renormalization.
Hence, the renormalization group evolution simplifies to 
\begin{equation}
  \label{eq:RGE_MStilde}
  \widetilde{\Pi}^S(X,\mu_1)=\exp\left(\;\int\limits_{\tilde{a}_s(\mu_0)}^{\tilde{a}_s(\mu_1)}\frac{\text{d}z}{z}\frac{2\gamma^S(z)}{\beta(z)}\right)\tilde{\Pi}^S(X,\mu_0)\,.
\end{equation}
The evolution equation for the $\MSbar$ scheme is obtained from
Eq. (\ref{eq:RGE_MStilde}) by simply replacing $\MStilde$ quantities by
their $\MSbar$ counterparts.

\subsection{The X-space scheme}
\label{sec:positionspace}

The X-space renormalization scheme is defined by fixing the correlator
of the normalized current  $j_X = Z^S_X j$
at a separation $X_0$ to its value in the free continuum theory \cite{Gimenez:2004me}:
\begin{equation}
  \label{eq:def_x}
  \Pi^S_X(X_0)=(Z^S_X)^2\,\Pi^S_0(X_0)=\Pi^S(X_0)\big|_\text{free}\,.
\end{equation}
This prescription can be readily implemented both in lattice and
perturbative QCD. In perturbation theory the free theory value of a
correlator is obviously just the leading order contribution.

\subsection{Other correlators}
\label{sec:v_t_corr}

In addition to scalar correlators, we also consider correlators of
vector, tensor, pseudo-scalar and axial-vector quark currents. In
position space these are defined as
\begin{align}
  \label{eq:def_corr}
   \Pi^V_{\mu\nu}(X)=&\langle j_\mu(X) j_\nu(0) \rangle\,,&\Pi^T_{\mu\nu\rho\sigma}(X)=&\langle j_{\mu\nu}(X) j_{\rho\sigma}(0) \rangle\,,\notag\\
   \Pi^P(X)=&\langle j_5(X) j_5(0)\rangle\,,& \Pi^A_{\mu\nu}(X)=&\langle j_{\mu
    5}(X) j_{\nu 5}(0)\rangle
\end{align}
with 
\begin{equation}
  \label{eq:def_j}
j_5=i\bar{\psi} \gamma_5 \psi,\quad j_\mu=\bar{\psi} \gamma_\mu \psi,\quad j_{\mu 5}=\bar{\psi} \gamma_\mu\gamma_5 \psi,\quad j_{\mu\nu}=\bar{\psi} \sigma_{\mu\nu} \psi\,.
\end{equation}
Since we work with $m_\psi=0$, the results for the pseudo-scalar
correlator will be the same as for the scalar correlator.

Except for two small points, the entire discussion for the scalar case also
holds for the more complicated Lorentz structures. First, in contrast to
all other correlators, the vector and the axial-vector correlators do 
receive contributions from diagrams with purely gluonic cuts. We choose
to neglect them in this work. This implies that also vector- and
 axial-vector correlators coincide. Second, it is not possible to
na\"ively renormalise the vector correlator according to the X-space
condition (Eq.~\eqref{eq:def_x}). The reason for this is that in
position space its tensor structure varies between different orders of
perturbation theory. We choose to renormalise the trace of the vector
correlator instead.

\section{Momentum space correlators: results}
\label{sec:q_space_res}

In the following two   sections   we present the results for the correlators both in momentum
and position space.
All results with their explicit renormalization scale dependence can also be retrieved from
\\{\tt http://www-ttp.particle.uni-karlsruhe.de/Progdata/ttp10/ttp10-42/}


\ice{ It is convenient to first compute the
correlators in momentum space and then apply Fourier transformation (see
Appendix \ref{sec:FT}) to obtain position space results.
}


\ice{
In momentum space, the correlators can be computed easily with the
FORM \cite{vermaseren-form} package MINCER \cite{Gorishnii:1989gt,Larin:1991fz} up to
three loops, i.e. order $\alpha_s^2$
The corresponding results for  

Moreover, we can also employ partial
higher order results. For dimensional reasons, only logarithmic terms
contribute to the Fourier transformation (for a more detailed
explanation see Appendix \ref{sec:FT}). These terms can in principle be extracted
from the corresponding $R$-ratios, which have
been computed in Refs. \cite{Baikov:2004ku,Baikov:2005rw} up to order
$\alpha_s^4$ for the scalar and the vector correlators. In practice it
is easier to reconstruct them using the renormalization group
equation~\eqref{eq:rge_log} and the anomalous dimensions, which arise as
intermediate results in the computation of the $R$-ratios. (... something
about the tensor correlator ...).
}

We list the correlators in momentum space at the scale $\mu^2=Q^2$,
where all logarithms vanish. Results for arbitrary values of $\mu$ can
be recovered by solving the renormalization group equation (i.e. by
using Eq. \eqref{eq:RGE_q} or \eqref{eq:rge_log}). Note that the
anomalous dimensions listed in Appendix \ref{sec:anom_dim} allow the
reconstruction of all logarithms at {\em one order higher}, i.e. at
order $\alpha_s^4$ for the vector and scalar correlators and at order
$\alpha_s^3$ for the tensor correlator. 

\begin{align}
\label{eq:sq_MS}
  &\Pi^{S}(Q)=-\frac{3}{4\pi^2}\,\*Q^2\*\left(1+\sum_{n=1}^{\infty}C^{(n),s}a_s^n\right)\,,\notag\\
  &C^{(1),s}=\frac{131}{24}-2\,\*\zeta_3\notag\,,\\
  &C^{(2),s}=\frac{17645}{288}-\frac{353}{12}\*\zeta_3-\frac{1}{8}\*\zeta_4+\frac{25}{6}\*\zeta_5+n_f\*\Big(-\frac{511}{216}+\frac{2}{3}\*\zeta_3\Big)\notag\,,\\
  &C^{(3),s}=\frac{215626549}{248832}-\frac{1789009}{3456}\*\zeta_3+\frac{1639}{32}\*\zeta_3^2-\frac{1645}{1152}\*\zeta_4+\frac{73565}{1728}\*\zeta_5\notag\\
 &\phantom{C^{(3),s}=}+\frac{325}{192}\*\zeta_6-\frac{665}{72}\*\zeta_7+n_f\*\Big(-\frac{26364175}{373248}+\frac{22769}{864}\*\zeta_3-\frac{5}{6}\*\zeta_3^2-\frac{53}{48}\*\zeta_4\notag\\
&\phantom{C^{(3),s}=}+\frac{1889}{432}\*\zeta_5\Big)+n_f^2\*\Big(\frac{499069}{559872}-\frac{157}{1296}\*\zeta_3+\frac{1}{48}\*\zeta_4-\frac{5}{18}\*\zeta_5\Big)
\,,
\end{align}

\begin{align}
\label{eq:sv_MS}
  &\Pi^{V}_{\mu\nu}(Q)=\frac{5}{12\pi^2}(-Q^2\delta_{\mu\nu}+Q_\mu Q_\nu) \left(1+\sum_{n=1}^{\infty}C^{(n),v}a_s^n\right)\,,\notag\\
  &C^{(1),v}=\frac{11}{4}-\frac{12}{5}\*\zeta_3\notag\,,\\
  &C^{(2),v}=\frac{41927}{1440}-\frac{829}{30}\*\zeta_3+5\*\zeta_5+n_f\*\Big(-\frac{3701}{2160}+\frac{19}{15}\*\zeta_3\Big)\notag\,,\\
  &C^{(3),v}=\frac{31431599}{69120}-\frac{624799}{1440}\*\zeta_3+\frac{99}{2}\*\zeta_3^2+\frac{11}{16}\*\zeta_4+\frac{349}{32}\*\zeta_5-\frac{133}{12}\*\zeta_7\notag\\
 &\phantom{C^{(3),v}=}+n_f\*\Big(-\frac{1863319}{34560}+\frac{174421}{4320}\*\zeta_3-\zeta_3^2-\frac{11}{48}\*\zeta_4+\frac{109}{18}\*\zeta_5\Big)\notag\\
 &\phantom{C^{(3),v}=}+n_f^2\*\Big(\frac{196513}{155520}-\frac{809}{1080}\*\zeta_3-\frac{1}{3}\*\zeta_5\Big)
\,,
\end{align}

\begin{align}
\label{eq:st_MS}
  &\Pi^{T}_{\mu\nu\rho\sigma}(Q)=Q^2\big(C^t\,T^{(1)}_{\mu\nu\rho\sigma}+
   D^t\, T^{(2)}_{\mu\nu\rho\sigma}(Q)\big)
  \,,\notag\\
&C^t=\frac{1}{12\pi^2}\Big(1+\sum_{n=1}^{\infty}C^{(n),t}a_s^n\Big)\,,\qquad D^t=-\frac{5}{12\pi^2}\Big(1+\sum_{n=1}^{\infty}D^{(n),t}a_s^n\Big)\,,\notag\\
  &C^{(1),t}=\frac{491}{72}-6\*\zeta_3\notag\,,\\
  &C^{(2),t}=\frac{556475}{7776}-\frac{2657}{36}\*\zeta_3+\frac{7}{24}\*\zeta_4+\frac{25}{2}\*\zeta_5+n_f\*\Big(-\frac{667}{162}+\frac{32}{9}\*\zeta_3\Big)\notag\,,\\
  &D^{(1),t}=\frac{593}{180}-\frac{12}{5}\*\zeta_3\notag\,,\\
  &D^{(2),t}=\frac{566777}{19440}-\frac{265}{9}\*\zeta_3+\frac{7}{60}\*\zeta_4+5\*\zeta_5+n_f\*\Big(-\frac{1333}{810}+\frac{64}{45}\*\zeta_3\Big)
  \end{align}
with
\begin{align}
  \label{eq:conv_q}
  &a_s = \frac{\alpha_s\big(\sqrt{Q^2}\big)}{\pi}\,,\qquad \zeta_n =
  \sum_{k=1}^\infty\frac{1}{k^n}\,,\notag\\
&T^{(1)}_{\mu\nu\rho\sigma}=\delta_{\mu\rho}\delta_{\nu\sigma}-\delta_{\mu\sigma}\delta_{\nu\rho}\,,\notag\\
&T^{(2)}_{\mu\nu\rho\sigma}(Q)=
\frac{Q_\mu Q_\rho}{Q^2}\delta_{\nu\sigma}
-\frac{Q_\mu Q_\sigma}{Q^2}\delta_{\nu\rho}
-\frac{Q_\nu Q_\rho}{Q^2}\delta_{\mu\sigma}
+\frac{Q_\nu Q_\sigma}{Q^2}\delta_{\mu\rho}\,,
\end{align}
and $n_f$ active quark flavours.

\section{Position space correlators: results}
\label{sec:x_space_res}

The position space results are obtained by four-dimensional Fourier
transformation (see Appendix \ref{sec:FT}) of their momentum space
counterparts. As it was discussed in Section \ref{sec:intro} the
position space
correlators are not sensitive (at $x^2 \not= 0$) to the constant
(non-logarithmic) contributions to the momentum-space ones.  Thus, the
knowledge of the full vector and scalar momentum space correlators at
order $\als^3$ plus the ${\cal O}(\als^4)$ anomalous dimensions
$\gamma^{SS}$ and $\gamma^{VV}$ from Appendix \ref{sec:anom_dim}
allows us to present below the vector and scalar correlators in the
position space at order $\als^4$.
Similarly, the use of Eq.~\re{eq:rge_log} and  the 
 ${\cal O}(\als^3)$ anomalous dimensions $\gamma^{(1),TT}$ and
 $\gamma^{(2),TT}$ computed by us 
(see eqs.  in  Appendix \ref{sec:anom_dim} ) result to the {\em full}  
${\cal O}(\als^3)$ results for the position space tensor correlator.

We present the
results in the $\MStilde$ scheme (see Eq. \eqref{eq:def_MStilde}) at the
scale  $\mu^2=\frac{1}{X^2}$
which correspond to $\MSbar$ results at the scale
$\mu^2=\frac{4}{X^2}e^{-2\gamma_E}$. Results at an arbitrary
scale $\mu$ can again be obtained with the use of the renormalization
group evolution (Eq. \eqref{eq:RGE_MStilde}). The correlators in position space read
\begin{align}
\label{eq:sx_MS}
&\widetilde{\Pi}^{S}(X)=\frac{3}{\pi^4
      \big(X^2\big)^3}\left(1+\sum_{n=1}^{\infty}\widetilde{C}^{(n),s}\tilde{a}_s^n\right)\,,\notag\\
&\widetilde{C}^{(1),s}=\frac{2}{3}\notag\,,\\
&\widetilde{C}^{(2),s}=\frac{817}{144}-\frac{39}{2}\*\zeta_3+n_f\*\Big(-\frac{23}{72}+\frac{2}{3}\*\zeta_3\Big)\notag\,,\\
&\widetilde{C}^{(3),s}=\frac{150353}{5184}-\frac{5125}{54}\*\zeta_3+\frac{815}{12}\*\zeta_5+n_f\*\Big(-\frac{13361}{3888}+\frac{3}{4}\*\zeta_3-\frac{5}{6}\*\zeta_4-\frac{25}{9}\*\zeta_5\Big)\notag\\
&\phantom{\widetilde{C}^{(3),s}=}+n_f^2\*\Big(-\frac{383}{11664}+\frac{8}{27}\*\zeta_3\Big)\,,\notag\\
&\widetilde{C}^{(4),s}=+\frac{22254833}{497664}-\frac{592067}{5184}\*\zeta_3+\frac{458425}{432}\*\zeta_3^2+\frac{265}{18}\*\zeta_4-\frac{607225}{864}\*\zeta_5\notag\\
 &\phantom{\widetilde{C}^{(4),s}=}-\frac{1375}{32}\*\zeta_6-\frac{178045}{768}\*\zeta_7+n_f\*\Big(-\frac{8775605}{373248}-\frac{392129}{5184}\*\zeta_3\notag\\
 &\phantom{\widetilde{C}^{(4),s}=}-\frac{955}{16}\*\zeta_3^2-\frac{6731}{576}\*\zeta_4+\frac{45695}{216}\*\zeta_5+\frac{2875}{288}\*\zeta_6+\frac{665}{72}\*\zeta_7\Big)\notag\\
 &\phantom{\widetilde{C}^{(4),s}=}+n_f^2\*\Big(-\frac{224695}{2239488}+\frac{11263}{1296}\*\zeta_3+\frac{5}{6}\*\zeta_3^2+\frac{25}{96}\*\zeta_4-\frac{6515}{432}\*\zeta_5\Big)\notag\\
 &\phantom{\widetilde{C}^{(4),s}=}+n_f^3\*\Big(\frac{6653}{559872}-\frac{173}{1296}\*\zeta_3+\frac{1}{144}\*\zeta_4+\frac{5}{18}\*\zeta_5\Big)
\,,\displaybreak[0]\\[1em]
 \label{eq:vx_MS}
 &\widetilde{\Pi}^{V}_{\mu\nu}(X)= \frac{6}{\pi^4
        \big(X^2\big)^3}\Bigg[\left(\frac{\delta_{\mu\nu}}{2}-\frac{X_{\mu}
          X_\nu}{X^2}\right)\*\widetilde{C}^v+\delta_{\mu\nu}\*\widetilde{D}^v\Bigg]\,,\notag\\
 & \widetilde{C}^v=1+\sum_{n=1}^{\infty}\widetilde{C}^{(n),v}\tilde{a}_s^n\,,
\qquad
\widetilde{D}^v=\sum_{n=0}^{\infty}\widetilde{D}^{(n),v}\tilde{a}_s^n\,,\notag\\
&\widetilde{C}^{(1),v}=1\notag\,,\\
&\widetilde{C}^{(2),v}=\frac{61}{6}-11\*\zeta_3+n_f\*\Big(-\frac{11}{18}+\frac{2}{3}\*\zeta_3\Big)\notag\,,\\
&\widetilde{C}^{(3),v}=\frac{7309}{48}-\frac{989}{6}\*\zeta_3+\frac{275}{6}\*\zeta_5+n_f\*\Big(-\frac{2617}{144}+\frac{47}{3}\*\zeta_3-\frac{25}{9}\*\zeta_5\Big)\notag\\
 &\phantom{\widetilde{C}^{(3),v}=}+n_f^2\*\Big(\frac{277}{648}-\frac{8}{27}\*\zeta_3\Big)\,,\notag\\
&\widetilde{C}^{(4),v}=\frac{57640705}{20736}-\frac{278401}{108}\*\zeta_3+\frac{5445}{8}\*\zeta_3^2-\frac{133705}{288}\*\zeta_5-\frac{7315}{48}\*\zeta_7\notag\\
& \phantom{\widetilde{C}^{(4),v}=}+n_f\*\Big(-\frac{10278875}{20736}+\frac{50705}{144}\*\zeta_3-55\*\zeta_3^2+\frac{65975}{432}\*\zeta_5+\frac{665}{72}\*\zeta_7\Big)\notag\\
& \phantom{\widetilde{C}^{(4),v}=}+n_f^2\*\Big(\frac{1554751}{62208}-\frac{10691}{864}\*\zeta_3+\frac{5}{6}\*\zeta_3^2-\frac{1315}{108}\*\zeta_5\Big)\notag\\
& \phantom{\widetilde{C}^{(4),v}=}+n_f^3\*\Big(-\frac{503}{1458}+\frac{2}{27}\*\zeta_3+\frac{5}{18}\*\zeta_5\Big)
\,,\notag\displaybreak[0]\\[1em]
&\widetilde{D}^{(0),v}=\widetilde{D}^{(1),v}=0\notag\,,\\
&\widetilde{D}^{(2),v}=-\frac{11}{24}+\frac{n_f}{36}\notag\,,\\
&\widetilde{D}^{(3),v}=-\frac{101}{9}+\frac{121}{12}\*\zeta_3+n_f\*\Big(\frac{587}{432}-\frac{11}{9}\*\zeta_3\Big)+n_f^2\*\Big(-\frac{1}{27}+\frac{1}{27}\*\zeta_3\Big)\,,\notag\\
&\widetilde{D}^{(4),v}=-\frac{616333}{2304}+\frac{1111}{4}\*\zeta_3-\frac{3025}{48}\*\zeta_5+n_f\*\Big(\frac{111409}{2304}-\frac{1607}{36}\*\zeta_3+\frac{275}{36}\*\zeta_5\Big)\notag\\
 &\phantom{\widetilde{D}^{(4),v}=}+n_f^2\*\Big(-\frac{54373}{20736}+\frac{79}{36}\*\zeta_3-\frac{25}{108}\*\zeta_5\Big)+n_f^3\*\Big(\frac{325}{7776}-\frac{5}{162}\*\zeta_3\Big)\,,\notag\\
\displaybreak[0]\\[1em]
 \label{eq:tx_MS}
&\widetilde{\Pi}^{T}_{\mu\nu\rho\sigma}(X)=-\frac{6}{\pi^4
       \big(X^2\big)^3}\left(\frac{1}{2}T^{(1)}_{\mu\nu\rho\sigma}-
  T^{(2)}_{\mu\nu\rho\sigma}(X)\right)\left(1+\sum_{n=1}^{\infty}\widetilde{C}^{(n),t}\tilde{a}_s^n\right)\,,\notag\\
&\widetilde{C}^{(1),t}=2\notag\,,\\
&\widetilde{C}^{(2),t}=\frac{5303}{432}-\frac{143}{18}\*\zeta_3+n_f\*\Big(-\frac{443}{648}+\frac{2}{3}\*\zeta_3\Big)\notag\,,\\
&\widetilde{C}^{(3),t}=+\frac{8439437}{46656}-\frac{77977}{486}\*\zeta_3+\frac{29}{54}\*\zeta_4+\frac{2395}{108}\*\zeta_5\notag\\
 &\phantom{\widetilde{C}^{(3),t}=}+n_f\*\Big(-\frac{246515}{11664}+\frac{1985}{108}\*\zeta_3+\frac{5}{18}\*\zeta_4-\frac{25}{9}\*\zeta_5\Big)\notag\\
 &\phantom{\widetilde{C}^{(3),t}=}+n_f^2\*\Big(\frac{18287}{34992}-\frac{4}{9}\*\zeta_3\Big)\,,
\end{align}
with
\begin{align}
  \label{eq:conv_x}
  &\tilde{a}_s = \frac{\tilde{\alpha}_s\big(1/\sqrt{X^2}\big)}{\pi}\,,\qquad \zeta_n =
  \sum_{k=1}^\infty\frac{1}{k^n}\,,\notag\\
&T^{(1)}_{\mu\nu\rho\sigma}=\delta_{\mu\rho}\delta_{\nu\sigma}-\delta_{\mu\sigma}\delta_{\nu\rho}\,,\notag\\
&T^{(2)}_{\mu\nu\rho\sigma}(X)=
\frac{X_\mu X_\rho}{X^2}\delta_{\nu\sigma}
-\frac{X_\mu X_\sigma}{X^2}\delta_{\nu\rho}
-\frac{X_\nu X_\rho}{X^2}\delta_{\mu\sigma}
+\frac{X_\nu X_\sigma}{X^2}\delta_{\mu\rho}\,,
\end{align}
and $n_f$ active quark flavours.

In numerical form for $n_f=3$ the results read
\begin{align}
\label{eq:num_x_mstilde_s}
  &\widetilde{\Pi}^{S}(X)=\frac{3}{\pi^4
      \big(X^2\big)^3}\big(1+0.67\,\tilde{a}_s-16.3\,\tilde{a}_s^2-31\,\tilde{a}_s^3+497\,\tilde{a}_s^4\big)\,,\\
\label{eq:num_x_mstilde_v}
  &\widetilde{\Pi}_{\mu\nu}^{V}(X)=\frac{6}{\pi^4
       \big(X^2\big)^3}\left(\frac{\delta_{\mu\nu}}{2}-\frac{X_{\mu}
          X_\nu}{X^2}\right)\big(1+\tilde{a}_s-2.5\,\tilde{a}_s^2-4.4\,\tilde{a}_s^3+66\,\tilde{a}_s^4\big)\notag\\
&\phantom{\widetilde{\Pi}_{\mu\nu}^{V}(X)=}-\delta_{\mu\nu}\big(-0.375\,\tilde{a}_s^2+0.63\,\tilde{a}_s^3+7.0\,\tilde{a}_s^4\big)\,,\\
\label{eq:num_x_mstilde_t}
  &\widetilde{\Pi}_{\mu\nu\rho\sigma}^{T}(X)=-\frac{6}{\pi^4
       \big(X^2\big)^3}\left(\frac{1}{2}T^{(1)}_{\mu\nu\rho\sigma}-
   T^{(2)}_{\mu\nu\rho\sigma}(X)\right)\big(1+2\,\tilde{a}_s+3.1\,\tilde{a}_s^2+6.6\,\tilde{a}_s^3\big)\,.
\end{align}
This can be compared to the $\MSbar$ results at the same scale $\mu^2=1/X^2$:
\begin{align}
\label{eq:num_x_msbar_s}
 &\Pi^{S}(X)= \frac{3}{\pi^4
      \big(X^2\big)^3}\big(1 + 0.20 \,a_s - 18.5 \,a_s^2 - 11
    \,a_s^3 + 579 \,a_s^4\big)\,,\\
\label{eq:num_x_msbar_v}
&\Pi_{\mu\nu}^{V}(X)= \frac{6}{\pi^4
  \big(X^2\big)^3}\left(\frac{\delta_{\mu\nu}}{2}-\frac{X_{\mu}
          X_\nu}{X^2}\right)\big(1+a_s-3.0\,a_s^2-2.4\,a_s^3+74\,a_s^4\big)\notag\\
&\phantom{\Pi_{\mu\nu}^{V}(X)=}-\delta_{\mu\nu}\big(-0.375\,a_s^2+1.0\,a_s^3+6.4\,a_s^4\big)\,,\\
\label{eq:num_x_msbar_t}
&\Pi_{\mu\nu\rho\sigma}^{T}(X)=-\frac{6}{\pi^4
       \big(X^2\big)^3}\left(\frac{1}{2}T^{(1)}_{\mu\nu\rho\sigma}-
  T^{(2)}_{\mu\nu\rho\sigma}(X)\right)\big(1. + 2.15\,a_s  + 3.3 \,a_s^2+6.4\,a_s^3\big)\,,
\end{align}
where we use the abbreviation
$a_s=\alpha_s\big(1/\sqrt{X^2}\big)/\pi$. In both schemes the
$\alpha_s^4$ coefficients of the vector and the scalar correlator are quite large. In the $\MStilde$ scheme we
observe a slightly smaller $\alpha_s^4$ contribution; furthermore the
coefficients are a bit more uniform between different orders. This
indicates a better behavior of the perturbative series in this scheme.

It is remarkable that the Lorentz structure of the tensor correlator remains
the same at each order while it begins to
vary at order $\alpha_s^2$ in the vector correlator result. 
This change can be easily understood by considering the transversality condition
\begin{equation}
  \label{eq:vx_trans}
  \partial_\mu \langle j_\mu(X) j_\nu(0) \rangle=0\,.
\end{equation}
Together with the appearance of terms which are logarithmic in $X$ also
the Lorentz structure has to change so that the vector correlator
remains transversal (see Eq.~\re{eq:FT_trans_1} in Appendix~\ref{sec:FT}).

As a consequence of this, it is not possible to renormalise the vector
correlator in the X-space scheme according to the prescription
Eq. \eqref{eq:def_x}. There are many possible generalizations leading
to a well-defined renormalization prescription for the vector correlator.
Among them, we choose to renormalise the trace
of the correlator. For the other correlators the renormalization is
straightforward. We refrain from presenting the somewhat lengthy
explicit results here. They can be easily constructed from the
$\MStilde$ results (Eq.~\eqref{eq:sx_MS}-\eqref{eq:tx_MS}) and the conversion formulas listed in 
Section~\ref{sec:conv} (Eq.~\eqref{eq:rc_s}-\eqref{eq:rc_t}).

\section{Bibliographical and technical comments}
\label{comments}

\newcommand{\info}{
\renewcommand{\arraystretch}{1.25}
\begin{table}[th]
\begin{center}
    \begin{tabular}{| l | l |  p{4.5cm} | p{3.7cm}  |}
    \hline

   & $\als^2$ & $\als^3$ & $\als^4$  
\\
\hline
$\Pi^{S}$ & \cite{Chetyrkin:1996sr,Gracey:2009da}, MINCER &  \cite{Baikov:2005rw,Baikov:prel},  $1/D$ expansion  &      
\\
                     &                 &                        &  
\\
\hline
$\gamma^{SS}$ & \cite{Gorishnii:1990zu,Chetyrkin:1996sr}, MINCER &  \cite{Chetyrkin:1996sr}, MINCER, IRR &  \cite{Baikov:2005rw,Baikov:prel}, $1/D$ expansion,
\\
                     &                 &                        &  \mbox{$R^*$-operation}
\\

\hline
$\text{Im}\, \Pi^{S}$ & \cite{Gorishnii:1990zu}, MINCER &  \cite{Chetyrkin:1996sr}, MINCER, \mbox{$R^*$-operation}  &  \cite{Baikov:2005rw}, $1/D$ expansion, 
\\
                     &                 &            &  \mbox{$R^*$-operation}
\\
\hline
$\Pi^{V}$ & \cite{GorKatLar91:R(s):4l,SurSam91,Gracey:2009da}, MINCER &  \cite{Baikov:2008jh,Baikov:2009uw},  $1/D$ expansion  &      
\\
                     &                 &                        &  
\\
\hline
$\gamma^{VV}$ & \cite{Chetyrkin:1979bj}, MINCER &  \cite{GorKatLar91:R(s):4l,SurSam91}, MINCER, IRR &  \cite{Baikov:2008jh,Baikov:2009uw}, $1/D$ expansion, 
\\
                     &                 &                        &  \mbox{$R^*$-operation}
\\
\hline
$\text{Im}\, \Pi^{V}$ & \cite{Chetyrkin:1979bj}, IRR &  \cite{Chetyrkin:1996sr}, MINCER, \mbox{$R^*$-operation}  &  \cite{Baikov:2009uw}, $1/D$ expansion,  
\\
                     &                 &            &  \mbox{$R^*$-operation}
\\
\hline
$\Pi^{T}$ & \cite{Gracey:2009da},  MINCER &               &      
\\
                     &                 &                        &  
\\
\hline
$\gamma^{TT}$ & \cite{Gracey:2009da}, MINCER &  present work,  MINCER, \mbox{$R^*$-operation} &   
\\
                     &                 &                        &  
\\
\hline
$\text{Im}\, \Pi^{T}$ & \cite{Gracey:2009da}, MINCER &  present work, MINCER, \mbox{$R^*$-operation}  &   
\\
\hline
    \end{tabular}
\end{center}
\vspace{3mm}
\caption{Calculations  of massless quark correlators in QCD: references and main theoretical tools.   
\label{table:info}
}
\end{table}
}

We start with some generic notes. First, the non-logarithmic
 part of a quark current correlator in momentum space
is {\em not physical}  within QCD as it requires  additional UV subtractions  
beyond the ones  associated to  the coupling constant renormalization.  

Still, the constant part {\em is} important because it provides us
with a very convenient way to compute the $Q$-dependent (that is  physical) 
contribution. Indeed, suppose that we want to get the $Q$-dependent 
contribution at order $\als^{n}$ to, say, the scalar correlator (the discussion 
below is  general and  applicable to every massless correlator). A direct calculation would require to deal with (n+1)-loop massless propagator-like 
diagrams contributing to $\Pi^S$. A better way\footnote{The observation  below
was explicitly made (for the particular case of the vector correlator) in \cite{Chetyrkin:1979bj}.}
is to use the evolution equation  \re{eq:rge_log}. Indeed, after a  (trivial) integration
of the right-hand side of the equation  with respect to $l_{\mu Q}$ we
arrive at the conclusion that
the $Q$-dependent part of   $\Pi^S$ is completely  determined by the  knowledge of three
ingredients:

1.  the  very correlator $\Pi^S$ ({\em including its constant, $Q$-independent part})
    and the beta-function to  order $\als^{n}$ (both are contributed by $n$-loop diagrams); 

2.  the  anomalous dimension    $\gamma^S$ to  order $\als^{n}$   ( $n$-loop diagrams);

3.  the  anomalous dimension $\gamma^{SS}$  to order  $\als^{n+1}$ ($(n+1)$-loop diagrams).

\info

Moreover, it is a well-known fact that the methods of Infrared Rearrangements (IRR) \cite{Vladimirov:IRR:80}
and $R^*$-operation \cite{Chetyrkin:1984xa} allow to reduce the problem of evaluation of a (n+1) loop 
contribution to an (arbitrary) anomalous dimension to the computation of some properly 
constructed set of  {\bf  n-loop} massless propagator-like diagrams.

Up to and including three loops the massless propagator-like diagrams
can be computed easily with the FORM \cite{vermaseren-form} package
MINCER \cite{Gorishnii:1989gt,Larin:1991fz}.  The package implements the algorithm
developed in \cite{Chetyrkin:1981qh} and is based on the use of the traditional method of
integration by parts.  Another powerful approach --- the method of
the Gegenbauer Polynomials in x-space (GPTx) \cite{Chetyrkin:1980pr} --- is less
automated, but, sometimes, is applicable in  cases with loop number
exceeding three  (see, e.g. \cite{Fiamberti:2009jw} for recent spectacular  examples).

The only systematical way to compute massless propagators at four
loops is based on the so-called $1/D$-expansion elaborated in
\cite{Baikov:2005nv,Baikov:2007zza} and   on the use of 
a special parametric representation of Feynman integrals
\cite{Baikov:tadpoles:96,Baikov:1996cd,Baikov:explit_solutions:97}. 
Here by computation we mean the reduction to the
corresponding {\em master} integrals, the latter have been 
computed analytically   \cite{Baikov:2010hf} and numerically  \cite{Smirnov:2010hd}.

In Table \ref{table:info} we display the bibliographic information
about results listed in Sections \ref{sec:q_space_res},\ref{sec:x_space_res}, and in Appendix B. In
addition we also mention the main theoretical tools used in obtaining these
results.

We want also to stress, that literally {\bf all} ${\cal
O}(\alpha_s^4)$ calculations  listed in the Table would be not possible
to perform without a heavy use of the parallel versions of FORM, PARFORM 
\cite{Fliegner:2000uy,Tentyukov:2004hz,Tentyukov:2006pr}
and TFORM \cite{Tentyukov:2007mu}. Last, but not least,  
the  generation of  thousands of four-and five-loop
input QCD diagrams  have  been conveniently done  with the help of 
the (FORTRAN) program QGRAF \cite{qgraf}.

\ice{

At present  there
The main method to compute Calculation of massless propga

In momentum space, the correlators can be computed easily with the

FORM \cite{vermaseren-form} package MINCER \cite{Larin:1991fz} up to
three loops, i.e. order $\alpha_s^2$
The main method to compute Calculation of massless propga
}

\section{Conversion between X-space scheme and $\MStilde$ or $\MSbar$}
\label{sec:conv}

The natural scale for a transition between the $\MStilde$ and
the X-space scheme is
\begin{equation}
  \label{eq:mutilde0}
  \tilde{\mu}_0^2=\frac{1}{X_0^2}\,.
\end{equation}
This is obviously equivalent to a transition between $\MSbar$ and
 X-space scheme at a scale $\mu_0^2= \frac{4}{X_0^2}e^{-2\gamma_E}$.
Imposing the X-space renormalization condition (Eq.~\eqref{eq:def_x}) on
\begin{equation}
  \label{eq:pix_pimsbar}
  \Pi^\delta_{X}(X_0)=\left(\frac{Z^\delta_X}{\widetilde{Z}^\delta(\tilde{\mu}_0)}\right)^2\widetilde{\Pi}^\delta(X_0,\tilde{\mu}_0)=\left(\frac{Z^\delta_X}{Z^\delta(\mu_0)}\right)^2\Pi^\delta(X_0,\mu_0)
\end{equation}
with $\delta \in \{V,\,S,\,T\}$ directly yields the desired ratios \cite{Gimenez:2004me} 
\begin{equation}
  \label{eq:rc_ratio}
  \frac{Z^\delta(\mu_0)}{Z_X^\delta}=\frac{\widetilde{Z}^\delta(\tilde{\mu}_0)}{Z_X^\delta}=  \sqrt{\frac{\widetilde{\Pi}^\delta(X_0,\tilde{\mu}_0)}{\Pi^\delta(X_0)\big|_\text{free}}}
\end{equation}
between the renormalization constants
in the different schemes. We obtain
\begin{align}
\label{eq:rc_s}
 &\frac{\widetilde{Z}^S(\tilde{\mu}_0)}{Z_X^S}=1+\sum_{n=1}^\infty
 \delta^{(n),s}\,\tilde{a}_s(\tilde{\mu}_0)^n\,,\notag\\
&\delta^{(1),s}=\frac{1}{3}\,,\notag\\
&\delta^{(2),s}=\frac{89}{32}-\frac{39}{4}\*\zeta_3+n_f\*\Big(-\frac{23}{144}+\frac{1}{3}\*\zeta_3\Big)\,,\notag\\
&\delta^{(3),s}=\frac{140741}{10368}-\frac{2387}{54}\*\zeta_3+\frac{815}{24}\*\zeta_5+n_f\*\Big(-\frac{12947}{7776}+\frac{19}{72}\*\zeta_3-\frac{5}{12}\*\zeta_4-\frac{25}{18}\*\zeta_5\Big)\notag\\
 &\phantom{\delta^{(3),s}=}+n_f^2\*\Big(-\frac{383}{23328}+\frac{4}{27}\*\zeta_3\Big)\,,\notag\\
&\delta^{(4),s}=\frac{13901515}{995328}-\frac{4393}{288}\*\zeta_3+\frac{208679}{432}\*\zeta_3^2+\frac{265}{36}\*\zeta_4-\frac{626785}{1728}\*\zeta_5\notag\\
 &\phantom{\delta^{(4),s}=}-\frac{1375}{64}\*\zeta_6-\frac{178045}{1536}\*\zeta_7+n_f\*\Big(-\frac{8029687}{746496}-\frac{418799}{10368}\*\zeta_3\notag\\
 &\phantom{\delta^{(4),s}=}-\frac{851}{32}\*\zeta_3^2-\frac{6571}{1152}\*\zeta_4+\frac{45895}{432}\*\zeta_5+\frac{2875}{576}\*\zeta_6+\frac{665}{144}\*\zeta_7\Big)\notag\\
 &\phantom{\delta^{(4),s}=}+n_f^2\*\Big(-\frac{257315}{4478976}+\frac{11273}{2592}\*\zeta_3+\frac{13}{36}\*\zeta_3^2+\frac{25}{192}\*\zeta_4-\frac{6515}{864}\*\zeta_5\Big)\notag\\
 &\phantom{\delta^{(4),s}=}+n_f^3\*\Big(\frac{6653}{1119744}-\frac{173}{2592}\*\zeta_3+\frac{1}{288}\*\zeta_4+\frac{5}{36}\*\zeta_5\Big)\,,\displaybreak[0]\\[1em]
\label{eq:rc_v}
 &\frac{\widetilde{Z}^V(\tilde{\mu}_0)}{Z_X^V}=(Z_X^V)^{-1}=1+\sum_{n=1}^\infty
 \delta^{(n),v}\,\tilde{a}_s(\tilde{\mu}_0)^n\,,\notag\\
&\delta^{(1),v}=\frac{1}{2}\,,\notag\\
&\delta^{(2),v}=\frac{97}{24}-\frac{11}{2}\*\zeta_3+n_f\*\Big(-\frac{1}{4}+\frac{1}{3}\*\zeta_3\Big)\,,\notag\\
&\delta^{(3),v}=\frac{14881}{288}-\frac{119}{2}\*\zeta_3+\frac{275}{12}\*\zeta_5+n_f\*\Big(-\frac{5395}{864}+\frac{47}{9}\*\zeta_3-\frac{25}{18}\*\zeta_5\Big)\notag\\
 &\phantom{\delta^{(3),v}=}+n_f^2\*\Big(\frac{181}{1296}-\frac{2}{27}\*\zeta_3\Big)\,,\notag\\
&\delta^{(4),v}=\frac{34042561}{41472}-\frac{294371}{432}\*\zeta_3+\frac{5203}{16}\*\zeta_3^2-\frac{212905}{576}\*\zeta_5-\frac{7315}{96}\*\zeta_7\notag\\
 &\phantom{\delta^{(4),v}=}+n_f\*\Big(-\frac{6096767}{41472}+\frac{7819}{96}\*\zeta_3-\frac{77}{3}\*\zeta_3^2+\frac{79775}{864}\*\zeta_5+\frac{665}{144}\*\zeta_7\Big)\notag\\
 &\phantom{\delta^{(4),v}=}+n_f^2\*\Big(\frac{889699}{124416}-\frac{2899}{1728}\*\zeta_3+\frac{13}{36}\*\zeta_3^2-\frac{1415}{216}\*\zeta_5\Big)\notag\\
 &\phantom{\delta^{(4),v}=}+n_f^3\*\Big(-\frac{1037}{11664}-\frac{2}{81}\*\zeta_3+\frac{5}{36}\*\zeta_5\Big)\,,\displaybreak[0]\\[1em]
\label{eq:rc_t}
 &\frac{\widetilde{Z}^T(\tilde{\mu}_0)}{Z_X^T}=1+\sum_{n=1}^\infty
 \delta^{(n),t}\,\tilde{a}_s(\tilde{\mu}_0)^n\,,\notag\\
&\delta^{(1),t}=1\,,\notag\\
&\delta^{(2),t}=\frac{4871}{864}-\frac{143}{36}\*\zeta_3+n_f\*\Big(-\frac{443}{1296}+\frac{1}{3}\*\zeta_3\Big)\,,\notag\\
&\delta^{(3),t}=\frac{7913369}{93312}-\frac{18529}{243}\*\zeta_3+\frac{29}{108}\*\zeta_4+\frac{2395}{216}\*\zeta_5\notag\\
 &\phantom{\delta^{(3),t}=}+n_f\*\Big(-\frac{238541}{23328}+\frac{1913}{216}\*\zeta_3+\frac{5}{36}\*\zeta_4-\frac{25}{18}\*\zeta_5\Big)+n_f^2\*\Big(\frac{18287}{69984}-\frac{2}{9}\*\zeta_3\Big)\,.
\end{align}
The conversion can of course be performed for an arbitrary scale $\mu$ with the
help of the renormalization group evolution (Eq. \eqref{eq:RGE_MStilde}).

\begin{figure}
  \centering
{\sf
  scalar correlator\\
    \includegraphics[]{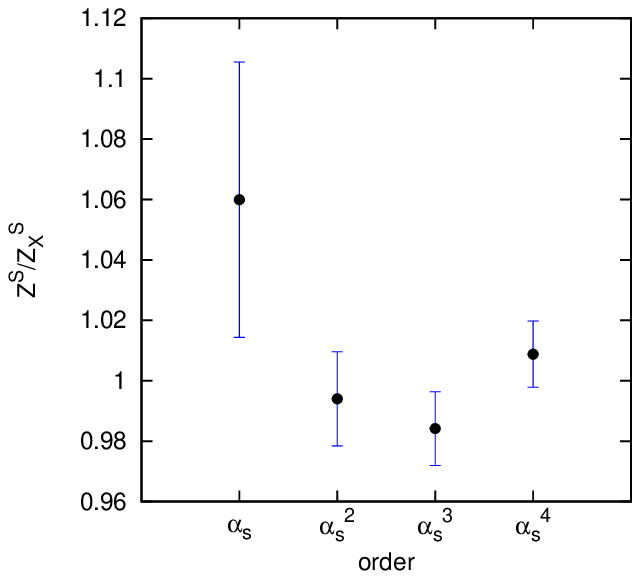}\includegraphics[]{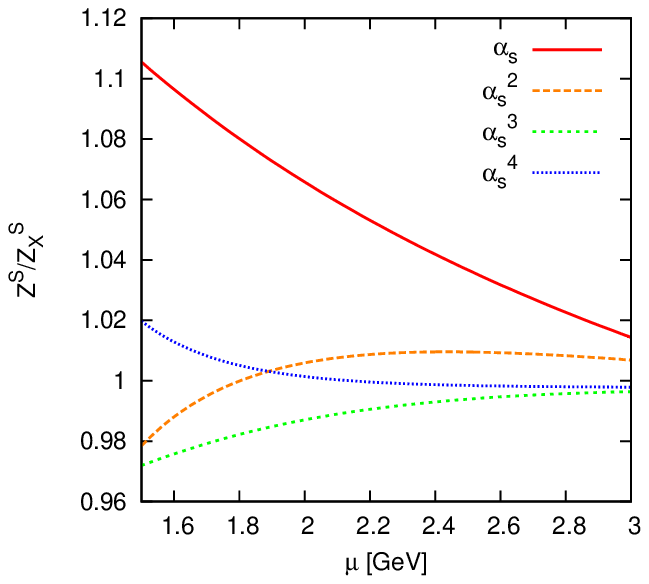}\\
vector correlator\\
    \includegraphics[]{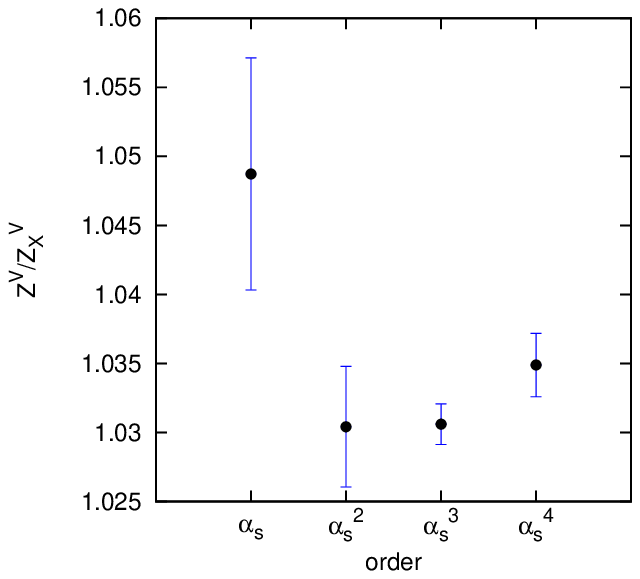}\includegraphics[]{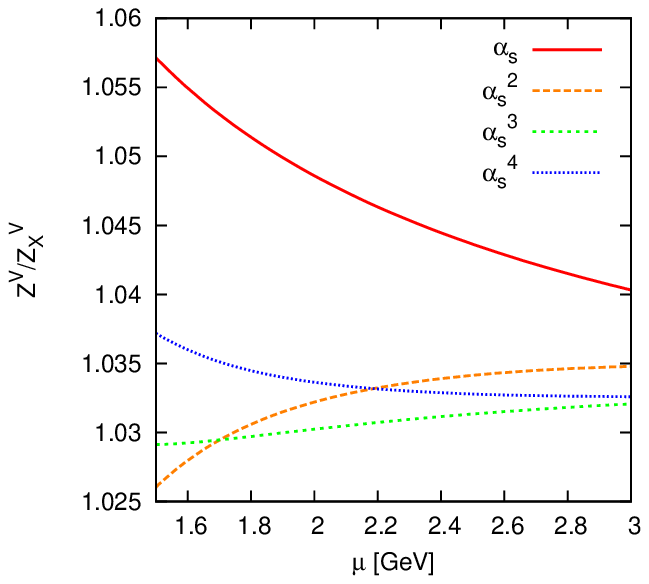}\\
  tensor correlator\\
    \includegraphics[]{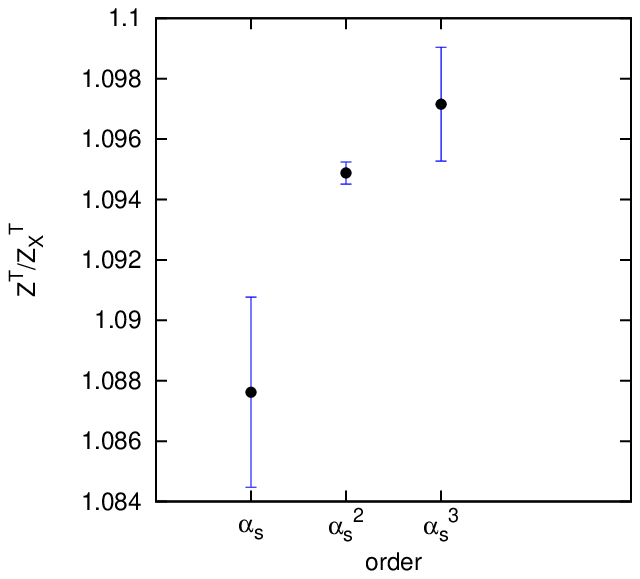}\includegraphics[]{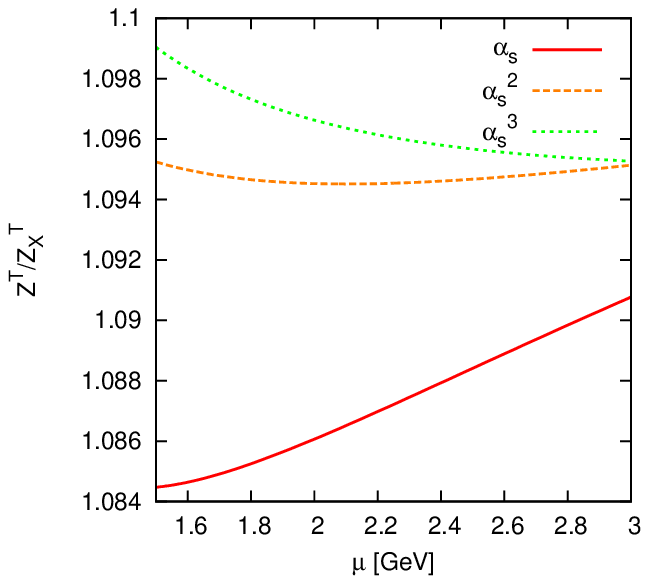}
}
\caption{Values of $Z^\delta(2\,\text{GeV})/Z_X^\delta(X_0^2)$  for
  vector, scalar and tensor currents with
    $X_0^2=(1.5\,\text{GeV})^{-2}$ and $n_f=3$
    quark flavours at different orders in perturbation
    theory. The error bars on the left-hand side are obtained by
    running the ratio of renormalization constants from  an intermediate
    scale $\mu$ with $1/X_0^2 \leq \mu^2 \leq 4/X_0^2$ to $2\,\text{GeV}$. On the
    right-hand side, the dependence on the choice of $\mu$ is shown.}
  \label{fig:zmsbar_nf3}
\end{figure}

\begin{figure}
  \centering 
  {\sf
  scalar correlator\\
  \includegraphics[]{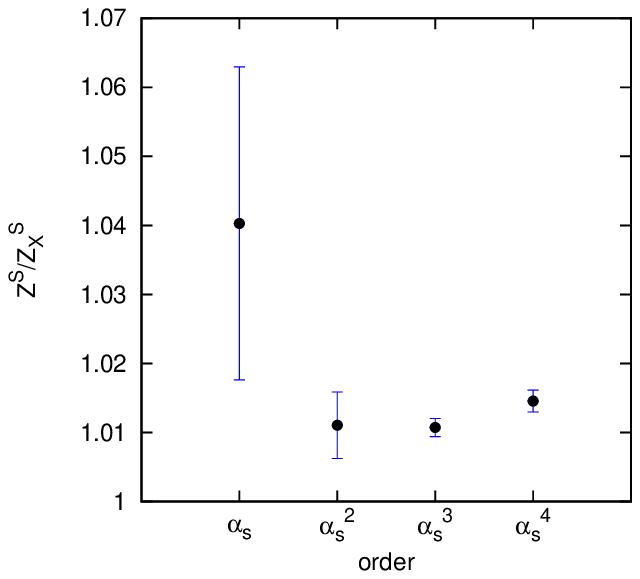}\includegraphics[]{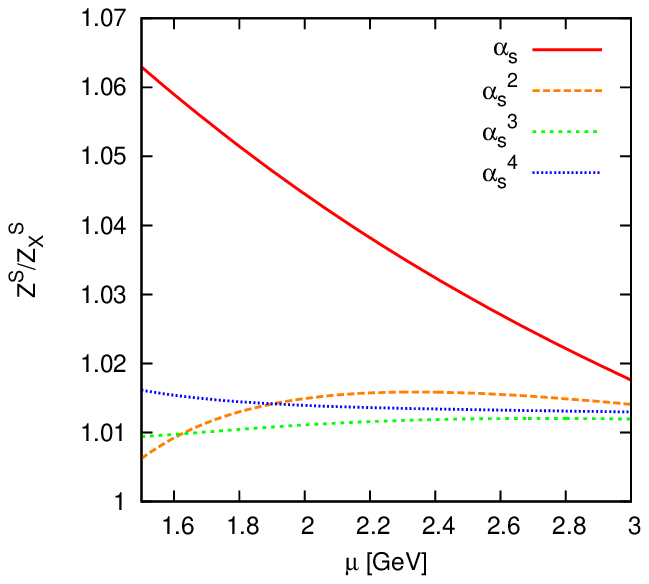}\\
  vector correlator\\
  \includegraphics[]{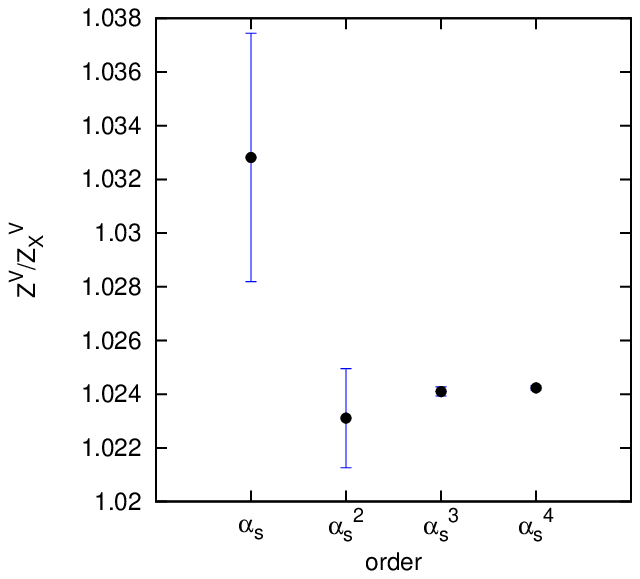}\includegraphics[]{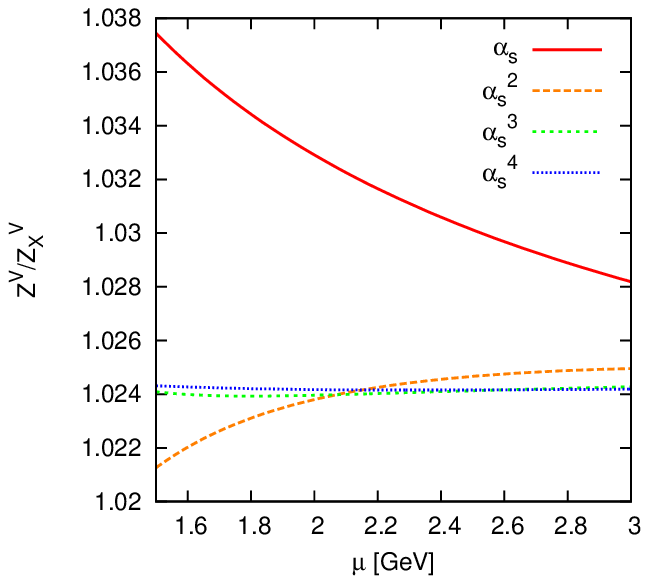}\\
  tensor correlator\\
  \includegraphics[]{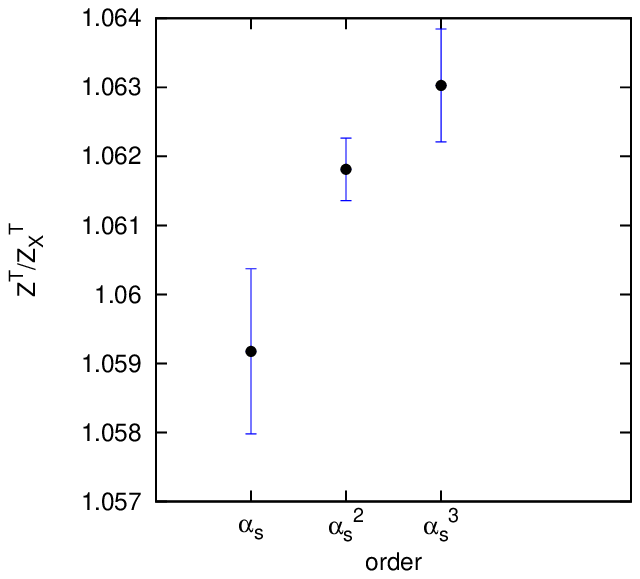}\includegraphics[]{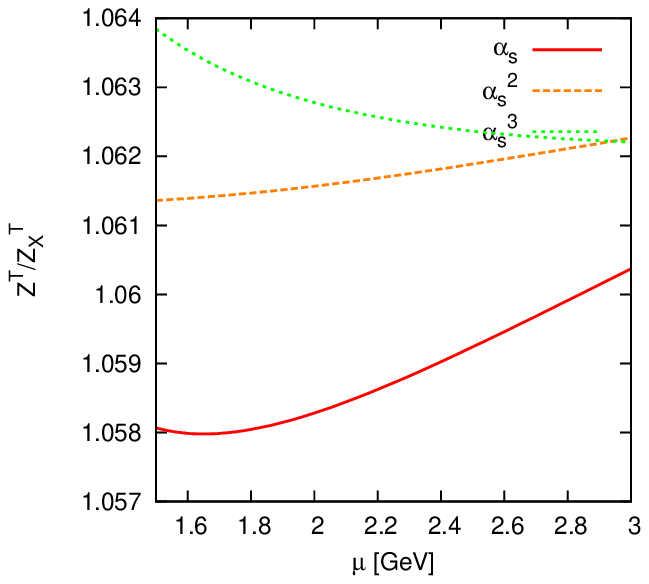}
}
\caption{Values of $Z^\delta(2\,\text{GeV})/Z_X^\delta(X_0^2)$  for
  vector, scalar and tensor currents with
    $X_0^2=(1.5\,\text{GeV})^{-2}$ and $n_f=0$
    quark flavours at different orders in perturbation
    theory.}
  \label{fig:zmsbar_nf0}
\end{figure}

The new higher order results reduce the theoretical error significantly
compared to the old NLO conversion formulas of
Ref. \cite{Gimenez:2004me}. As an example we show the values of $Z^\delta(2\,\text{GeV})/Z_X^\delta(X_0^2)$
with $\delta \in \{V,\,S,\,T\}$, $X_0^2=(1.5\,\text{GeV})^{-2}$ at different orders in perturbation theory. We estimate the
theory error from higher orders by performing the transition between the two
schemes at some varying intermediate scale $\mu$ and evolving the result
to the final scale of $2\,\text{GeV}$:
\begin{equation}
  \label{eq:RGE_z}
  \frac{Z^\delta(2\,\text{GeV})}{Z_X^\delta(X_0^2)} = \exp\left(\int\limits_{a_s(\mu)}^{a_s(2\,\text{GeV})}\frac{\text{d}z}{z}\frac{\gamma^\delta(z)}{\beta(z)}\right)\frac{Z^\delta(\mu)}{Z_X^\delta(X_0^2)}\,.
\end{equation}
For the evolution, we use anomalous dimensions at one order higher than
$Z^\delta(2\,\text{GeV})/Z_X^\delta(X_0^2)$, up to the highest available
order of $\alpha_s^4$. In Figure \ref{fig:zmsbar_nf3}
the values for this transition factor are plotted for a varying intermediate
scale $\mu^2$ between $1/X_0^2$ and $4/X_0^2$ and $n_f=3$ quark flavours.
Figure \ref{fig:zmsbar_nf0} shows the same ratios of renormalization
constants for $n_f=0$. As expected from the numerical formulas 
Eqs.~\eqref{eq:num_x_msbar_s}-\eqref{eq:num_x_msbar_t}
the vector and scalar correlators receive large contributions at order
$\alpha_s^2$ and $\alpha_s^4$ for $n_f=3$. 

\section{Conclusion}
\label{sec:conc}

In this work we have presented presently  available  information
on three basic quark currents  correlators  --- 
the scalar, vector and  the tensor ones --- considered within
the  massless QCD. The  correlators and  
the corresponding RG evolution equations have been studied both in the momentum and
position space. Explicit conversion formulas  relating
the $\MSbar$ renormalized vector, scalar and tensor currents to their counterparts
renormalized in the X-space renormalization scheme are constructed. 
It is demonstrated that the  new higher order results reduce the theoretical error significantly
compared to the old NLO conversion formulas of Ref. \cite{Gimenez:2004me}.

\section*{Acknowledgments}
\label{ack}

We thank P. ~Baikov and J. K\"uhn for attentive reading of the
manuscript and  for their kind permission to use in our work the
unpublished results related to evaluation of the scalar correlator.

This work was supported by the Deutsche Forschungsgemeinschaft in the
Sonderforschungsbereich/Transregio SFB/TR-9 ``Computational Particle
Physics''. A. M. thanks the Graduiertenkolleg ``Hochenergiephysik und
Teilchenastrophysik'' and the Landesgraduiertenf\"orderung des Landes
Baden--W\"urttemberg for support.

\appendix

\section{Euclidean correlators from   Minkowskian  ones}
\label{sec:min_to_eucl}

Let us consider a generic correlator defined originally in the  Minkowskian 
space:

\begin{align}
  \label{generic:M}
\Pi_{{\mu_1\mu_2 \dots} {\nu_1\nu_2 \dots }}(q) =&\,
i \int_{M} \text{d}x\ e^{iqx}\langle T j_{\nu_1\nu_2 \dots} (x) \,\, 
j^{\dagger}_{\mu_1\mu_2 \dots} (0)\rangle  
\\
=&\,\sum_k T^k_{{\mu_1\mu_2 \dots}(q) {\nu_1\nu_2 \dots }}(q)\, \Pi_k(-q^2)
{},
\end{align}
where the tensors $T^k_{{\mu_1\mu_2 \dots} {\nu_1\nu_2 \dots }}$ are made from 
the metric tensor $g_{\mu\nu}$, 
the vector $q$ and the indexes
${\mu_1\mu_2 \dots} {\nu_1\nu_2 \dots }$. 

The corresponding Euclidean correlator in the momentum space is constructed as 
follows:
\begin{equation}
  \label{generic:E}
\Pi_{{\mu_1\mu_2 \dots} {\nu_1\nu_2 \dots }}(Q) \equiv \sum_k T^k_{{\mu_1\mu_2 
\dots} {\nu_1\nu_2 \dots }}(Q) 
\,\,
\Pi_k(Q^2)
{},
\end{equation}
with 
$T^k_{{\mu_1\mu_2 \dots} {\nu_1\nu_2 \dots }}(Q)$ being made from 
$T^k_{{\mu_1\mu_2 \dots} {\nu_1\nu_2 \dots }}(q)$ with the help of 
the replacements 
$ q_{\mu_i} \to -Q_{\mu_i}, q_{\nu_i} \to -Q_{\nu_i}, g_{\mu\nu} \to -
\delta_{\mu\nu}$.

At last, the Euclidean correlator in the position space is {\em defined} with 
the  help of
the Fourier transformation, viz. 
\begin{equation}
  \label{generic:EX}
\Pi_{{\mu_1\mu_2 \dots} {\nu_1\nu_2 \dots }}(Q) \equiv
 \int_{E} \text{d}X\ e^{iQX}\langle T j_{\nu_1\nu_2 \dots }(X) \,\, 
j^{\dagger}_{\mu_1\mu_2 \dots} (0)\rangle  
\end{equation}

  \section{Fourier transformation in $\mathbf d$ dimensions}
  \label{sec:FT}

  Perturbative calculations are in most cases quite cumbersome in
  position space. A more convenient alternative is to perform the
  calculation in momentum space, where exploiting invariance under
  translations (implying momentum conservation) leads to great
  simplifications. In the end, the result in position space can be recovered by means of Fourier transformation.

\ice{
  This procedure is, however, less straightforward than it might seem at
  first glance: Green functions in quantum field theory are usually not
  defined as functions in the classical sense, but only as
  distributions. Bare Green functions are conveniently calculated in
  dimensional regularization; to transform them to position space in a
  consistent way it is necessary to generalize Fourier transformation to
  $d$ dimensions. Still, we cannot expect to obtain exactly the same
  results as from a genuine $x$-space calculation. First, additional
  {\it contact terms} originating from the limit $x\to0$ can
  occur. These contributions do not appear in the problem at hand, so we
  refrain from discussing them in this work. Second, the overall
  normalization of the respective bare Green functions can be different
  for $d\neq4$. This is not really a problem since the final results
  after renormalization and considering the limit $d\to4$ will not be
  affected.
}
\subsection{Fourier transformation of bare functions}
\label{sec:FT_bare}

  In the current work, all bare Green functions have a power-like
  dependence on the momentum. For the transformation from euclidean
  momentum space to euclidean position space we use the following
  formula:
  \begin{equation}
    \label{eq:FT}
    \FT\big((Q^2)^{-r}\big)\equiv\int \frac{\text{d}^dQ}{(2\pi)^d} \frac{e^{i Q
        X}}{(Q^2)^r} = \frac{1}{(4\pi)^\frac{d}{2}}
     \frac{\Gamma(\frac{d}{2}-r)}{\Gamma(r)}\left(\frac{X^2}{4}\right)^{r-\frac{d}{2}}\,.
  \end{equation}
  It is straightforward to derive this formula by using Schwinger
  parametrization and Gauss integration.

  For Green functions with a non-trivial Lorentz structure it is convenient
  to use momentum operators, i.e.
  \begin{align}
    \label{eq:FT_qmu}
    Q_\mu\to& - i \frac{\partial}{\partial X_\mu}\,,\\
    \label{eq:qmu_qnu}
    \FT\big(Q_\mu Q_\nu
    (Q^2)^{-(r+1)}\big)=&\frac{1}{(4\pi)^\frac{d}{2}}
    \frac{\Gamma(\frac{d}{2}-r)}{\Gamma(r)}\left(\frac{X^2}{4}\right)^{r-\frac{d}{2}}
    \left[\frac{\delta_{\mu\nu}}{2r}-\left(\frac{d}{2r}-1\right)\frac{X_\mu
        X_\nu}{X^2}\right]\,.
  \end{align}
  Note that forming the trace of the right-hand side of
  Eq. \eqref{eq:qmu_qnu} again yields the simpler result from
  Eq. \eqref{eq:FT}.

  We also see explicitly that the Fourier transformation preserves
  transversality. Combining Eqs. \eqref{eq:FT} and \eqref{eq:qmu_qnu} we
  obtain
  \begin{align}
    \label{eq:FT_trans_1}
    \FT\left[\left(\delta_{\mu\nu}-\frac{Q_\mu
          Q_\nu}{Q^2}\right)(Q^2)^{-r}\right]=&\frac{1}{(4\pi)^\frac{d}{2}}
    \frac{\Gamma(\frac{d}{2}-r)}{\Gamma(r)}
    \left(\frac{X^2}{4}\right)^{r-\frac{d}{2}}
\\
    &\quad \times \left[\left(1-\frac{1}{2r}\right)\delta_{\mu\nu}-\left(1-\frac{d}{2r}\right)\frac{X_\mu X_\nu}{X^2}\right]\,,\notag\\
    \label{eq:FT_trans_2}
    \partial_\mu \FT\left[\left(\delta_{\mu\nu}-\frac{Q_\mu
          Q_\nu}{Q^2}\right)(Q^2)^{-r}\right]=&\partial_\nu
    \FT\left[\left(\delta_{\mu\nu}-\frac{Q_\mu
          Q_\nu}{Q^2}\right)(Q^2)^{-r}\right] = 0\,.
  \end{align}

\subsection{Fourier transformation in four dimensions}
\label{sec:FT_4}

The Fourier transformation for the correlators we consider is regular at
$d=4$. This means that there are two ways of obtaining the renormalized
correlators in position space. On the one hand, we can transform the
bare Green functions in $d$ dimensions and perform the renormalization
afterwards. On the other hand, we can start from the renormalized
correlator in momentum space and transform it in four dimensions.
In the latter case we have to transform
terms which are logarithmic in $Q^2$ in addition to simple powers of
$Q^2$. Rewriting the logarithms with the help of derivatives leads to
\begin{equation}
  \label{eq:FT_log}
  \FT\big[\left(Q^2\right)^{r}\log^s\left(Q^2\right)\big]=\lim_{\delta \to
    0}\bigg(\frac{\partial}{\partial \delta}\bigg)^s \FT\big[\left(Q^2\right)^{r+\delta}\big]\,.
\end{equation}
In Table \ref{tab:FT} the resulting expressions are listed for all
powers of $Q^2$ and $\log\left(Q^2\right)$ which appear in the momentum
space correlators up to order $\alpha_s^4$.

  \begin{table}
    \centering
    \begin{tabular}{cl}
      \hline
      Momentum space & Position space\\
      \hline
$\log(\mu^2/Q^2)$&$\frac{1}{\pi^2\*(X^2)^2}$ \\
$\log^2(\mu^2/Q^2)$&$\frac{2}{\pi^2\*(X^2)^2}\*(-1+l_{\mu X})$ \\
$\log^3(\mu^2/Q^2)$&$\frac{3}{\pi^2\*(X^2)^2}\*(-2\*l_{\mu X}+l_{\mu X}^2)$ \\
$\log^4(\mu^2/Q^2)$&$\frac{4}{\pi^2\*(X^2)^2}\*(4\*\zeta_3-3\*l_{\mu X}^2+l_{\mu X}^3)$ \\
$\log^5(\mu^2/Q^2)$&$\frac{5}{\pi^2\*(X^2)^2}\*(-16\*\zeta_3+16\*l_{\mu X}\*\zeta_3-4\*l_{\mu X}^3+l_{\mu X}^4)$ \\
$Q^2\*\log(\mu^2/Q^2)$&$-\frac{8}{\pi^2\*(X^2)^3}$ \\
$Q^2\*\log^2(\mu^2/Q^2)$&$\frac{8}{\pi^2\*(X^2)^3}\*(5-2\*l_{\mu X})$ \\
$Q^2\*\log^3(\mu^2/Q^2)$&$\frac{24}{\pi^2\*(X^2)^3}\*(-4+5\*l_{\mu X}-l_{\mu X}^2)$ \\
$Q^2\*\log^4(\mu^2/Q^2)$&$\frac{16}{\pi^2\*(X^2)^3}\*(6-8\*\zeta_3-24\*l_{\mu X}+15\*l_{\mu X}^2-2\*l_{\mu X}^3)$ \\
$Q^2\*\log^5(\mu^2/Q^2)$&$\frac{40}{\pi^2\*(X^2)^3}\*(40\*\zeta_3+12\*l_{\mu
  X}-16\*l_{\mu X}\*\zeta_3-24\*l_{\mu X}^2+10\*l_{\mu X}^3-l_{\mu
  X}^4)$ \\[.3em]
\hline\\[-.7em]
    \end{tabular}
    \caption{Fourier transformation of logarithmic terms for various
      powers of $Q^2$ and $\log Q^2$. Terms that do not contain
      logarithms vanish for non-negative integer powers of $Q^2$. We use
    the abbreviation $l_{\mu X}=\log(\mu^2X^2/4)+2\gamma_E$.}
    \label{tab:FT}
\end{table}

\subsection{Logarithmic structure}
\label{sec:FT_log}

One-scale problems, like the correlators considered in this work,
exhibit a very special structure of their logarithmic terms both in
momentum and in position space.

In momentum space in the $\MSbar$ scheme, logarithms have the form
$\log(\mu^2/Q^2)$. They originate from terms of the form
\begin{equation}
  \label{eq:log_Q}
  \left(\frac{\mu^2}{Q^2}\right)^{l\epsilon}=1+l\epsilon\log\left(\frac{\mu^2}{Q^2}\right)+{\cal O}\big(\epsilon^2\big)\,,
\end{equation}
where $l$ denotes the number of loops in the corresponding diagram. In
order to obtain the logarithmic structure in position space we can again
extract terms of the form $y^{l\epsilon}$ from the Fourier transform of
the left-hand side of Eq.~\eqref{eq:log_Q}. 
Expressing the Gamma functions in the Fourier transform in terms of exponential functions and
polynomials in $\epsilon$ we find  
\begin{equation}
  \label{eq:log_X}
  \left(\frac{\mu^2X^2}{4}e^{2\gamma_E}\right)^{l\epsilon}=1+l\epsilon\Bigg[\log\left(\frac{\mu^2X^2}{4}\right)+2\gamma_E\Bigg]+{\cal O}\big(\epsilon^2\big)\,,
\end{equation}
for the general structure of position space logarithms.

\section{Anomalous dimensions}
\label{sec:anom_dim}
In our conventions the momentum space renormalization group equations of
the scalar, vector and tensor correlators read
\begin{align}
  \label{eq:RGE_all}
  &\mu^2\frac{\text{d}}{\text{d}\mu^2}\Pi^S(Q,\mu)=2\gamma^S\Pi^S(Q,\mu)+\gamma^{SS}Q^2\,,\\
  &\mu^2\frac{\text{d}}{\text{d}\mu^2}\Pi^V_{\mu\nu}(Q,\mu)=2\gamma^V\Pi^V_{\mu\nu}(Q,\mu)+\gamma^{VV}(-Q^2\delta_{\mu\nu}+Q_\mu
  Q_\nu)\,,\\
  &\mu^2\frac{\text{d}}{\text{d}\mu^2}\Pi^T_{\mu\nu\rho\sigma}(Q,\mu)=2\gamma^T\Pi^T_{\mu\nu\rho\sigma}(Q,\mu)+\big(\gamma^{(1),TT}T^{(1)}_{\mu\nu\rho\sigma}+\gamma^{(2),TT}T^{(2)}_{\mu\nu\rho\sigma}(Q)\big)Q^2
\end{align}
with the tensor structures
\begin{align}
  \label{eq:T1_T2}
&T^{(1)}_{\mu\nu\rho\sigma}=\delta_{\mu\rho}\delta_{\nu\sigma}-\delta_{\mu\sigma}\delta_{\nu\rho}\,,\notag\\
&T^{(2)}_{\mu\nu\rho\sigma}(Q)=
\frac{Q_\mu Q_\rho}{Q^2}\delta_{\nu\sigma}
-\frac{Q_\mu Q_\sigma}{Q^2}\delta_{\nu\rho}
-\frac{Q_\nu Q_\rho}{Q^2}\delta_{\mu\sigma}
+\frac{Q_\nu Q_\sigma}{Q^2}\delta_{\mu\rho}\,.
\end{align}
The corresponding evolution equations in position space are analogous,
but contain no subtractive anomalous dimensions.
The anomalous dimensions are given by
\begin{align}
  \label{eq:anom_dim}
&  \gamma^V(a_s)=\,0\,,\displaybreak[0]\\[1em]
&  \gamma^{VV}(a_s)=\frac{1}{16\pi^2}\sum_{n=0}^{\infty} \gamma^{VV}_na_s^n\,,\notag\\
&  \gamma^{VV}_0=4\,,\notag\\
&  \gamma^{VV}_1=4\,,\notag\\
&  \gamma^{VV}_2=\frac{125}{12}-\frac{11}{18} n_f\,,\notag\\
&  \gamma^{VV}_3=\frac{10487}{432}+\frac{110}{9} \zeta_{3}+ n_f\Big(-\frac{707}{216}-\frac{110}{27}  \,\zeta_{3}\Big)-\frac{77}{972}n_f^2\,,\notag\\
&  \gamma^{VV}_4=  \frac{2665349}{41472}+\frac{182335}{864}
\zeta_{3}-\frac{605}{16}  \zeta_{4}-\frac{31375}{288}  \zeta_{5}\notag\\
&  \phantom{\gamma^{VV}_4=}+  n_f\Big(-\frac{11785}{648}-\frac{58625}{864}  \zeta_{3}+\frac{715}{48}  \zeta_{4}+\frac{13325}{432}  \zeta_{5}\Big)\notag\\
&  \phantom{\gamma^{VV}_4=}+  n_f^2\Big(-\frac{4729}{31104}+\frac{3163}{1296}  \zeta_{3}-\frac{55}{72}  \zeta_{4}\Big)+n_f^3\Big(\frac{107}{15552}+\frac{1}{108}  \zeta_{3}\Big)\,,\displaybreak[0]\\[1em]
&  \gamma^S(a_s)=-\gamma_m=\sum_{n=0}^{\infty} \gamma^{S}_na_s^{n+1}\,,\notag\\
&  \gamma^S_0=1\,,\notag\\
&  \gamma^S_1=\frac{101}{24}-\frac{5}{36}\*n_f\,,\notag\\
&  \gamma^S_2=\frac{1249}{64}+n_f\*\Big(-\frac{277}{216}-\frac{5}{6}\*\zeta_3\Big)-\frac{35}{1296}\*n_f^2\,,\notag\\
&  \gamma^S_3=\frac{4603055}{41472}+\frac{530}{27}\*\zeta_3-\frac{275}{8}\*\zeta_5+n_f\*\Big(-\frac{91723}{6912}-\frac{2137}{144}\*\zeta_3+\frac{55}{16}\*\zeta_4+\frac{575}{72}\*\zeta_5\Big)\notag\\
 &\phantom{\gamma^S_3=}+n_f^2\*\Big(\frac{2621}{31104}+\frac{25}{72}\*\zeta_3-\frac{5}{24}\*\zeta_4\Big)+n_f^3\*\Big(-\frac{83}{15552}+\frac{1}{108}\*\zeta_3\Big)\,,\displaybreak[0]\\[1em]
&  \gamma^{SS}(a_s)=\frac{1}{16\pi^2}\sum_{n=0}^{\infty} \gamma^{SS}_na_s^n\,,\notag\\
&  \gamma^{SS}_0=-6\,,\notag\\
&  \gamma^{SS}_1=-10\,,\notag\\
&  \gamma^{SS}_2=-\frac{455}{12}+3\*\zeta_3-2\,n_f\,,\notag\\
&  \gamma^{SS}_3=-\frac{157697}{864}+\frac{1645}{36}\*\zeta_3-\frac{45}{4}\*\zeta_4-\frac{65}{2}\*\zeta_5+n_f\*\Big(\frac{14131}{1296}+\frac{26}{3}\*\zeta_3+\frac{11}{2}\*\zeta_4\Big)\notag\\
 &\phantom{\gamma^{SS}_3=}+n_f^2\*\Big(\frac{1625}{1944}-\frac{2}{3}\*\zeta_3\Big)
 \,,\notag\\
&  \gamma^{SS}_4=-\frac{1305623}{864}-\frac{540883}{3456}\*\zeta_3-\frac{19327}{288}\*\zeta_3^2-\frac{113557}{384}\*\zeta_4+\frac{158765}{576}\*\zeta_5+\frac{29825}{64}\*\zeta_6\notag\\
 &\phantom{\gamma^{SS}_4=}+\frac{97895}{384}\*\zeta_7+n_f\*\Big(\frac{11341807}{62208}+\frac{385147}{1728}\*\zeta_3-\frac{187}{16}\*\zeta_3^2+\frac{10207}{192}\*\zeta_4\notag\\
 &\phantom{\gamma^{SS}_4=}-\frac{55127}{288}\*\zeta_5-\frac{6725}{96}\*\zeta_6\Big)+n_f^2\*\Big(\frac{249113}{373248}-\frac{749}{48}\*\zeta_3+\frac{21}{8}\*\zeta_4+\frac{37}{4}\*\zeta_5\Big)\notag\\
 &\phantom{\gamma^{SS}_4=}+n_f^3\*\Big(\frac{1625}{15552}+\frac{5}{108}\*\zeta_3-\frac{1}{6}\*\zeta_4\Big)
\,,\displaybreak[0]\\[1em]
&  \gamma^T(a_s)=\sum_{n=0}^{\infty} \gamma^{T}_na_s^{n+1}\,,\notag\\
&  \gamma^T_0=-\frac{1}{3}\,,\notag\\
&  \gamma^T_1=-\frac{181}{72}+\frac{13}{108}\*n_f\,,\notag\\
&
\gamma^T_2=-\frac{52555}{5184}+\frac{29}{54}\*\zeta_3+n_f\*\Big(\frac{655}{648}+\frac{5}{18}\*\zeta_3\Big)+\frac{n_f^2}{144}\,,\notag\\
&  \gamma^T_3=-\frac{2208517}{41472}+\frac{7733}{3888}\*\zeta_3-\frac{319}{144}\*\zeta_4+\frac{10465}{972}\*\zeta_5\notag\\
 &\phantom{\gamma^T_3=}+n_f\*\Big(\frac{1537379}{186624}+\frac{18979}{3888}\*\zeta_3-\frac{437}{432}\*\zeta_4-\frac{575}{216}\*\zeta_5\Big)\notag\\
 &\phantom{\gamma^T_3=}+n_f^2\*\Big(-\frac{9961}{93312}-\frac{115}{648}\*\zeta_3+\frac{5}{72}\*\zeta_4\Big)
+n_f^3\*\Big(-\frac{7}{15552}-\frac{1}{324}\*\zeta_3\Big)
\,,\displaybreak[0]\\[1em]
&  \gamma^{(1),TT}(a_s)=\frac{1}{16\pi^2}\sum_{n=0}^{\infty} \gamma^{(1),TT}_na_s^n\,,\notag\\
&  \gamma^{(1),TT}_0=2\,,\notag\\
&  \gamma^{(1),TT}_1=\frac{22}{9}\,,\notag\\
&  \gamma^{(1),TT}_2=\frac{841}{324}+\frac{7}{9}\*\zeta_3+\frac{n_f}{81}\,,\notag\\
&  \gamma^{(1),TT}_3=\frac{617299}{69984}+\frac{35171}{972}\*\zeta_3-\frac{29}{36}\*\zeta_4-\frac{1955}{54}\*\zeta_5\notag\\
&\phantom{\gamma^{(1),TT}_3=}+n_f\*\Big(\frac{32821}{34992}-\frac{152}{81}\*\zeta_3+\frac{37}{54}\*\zeta_4\Big)+n_f^2\*\Big(\frac{557}{17496}-\frac{2}{27}\*\zeta_3\Big)\,,\displaybreak[0]\\[1em]
&  \gamma^{(2),TT}(a_s)=\frac{1}{16\pi^2}\sum_{n=0}^{\infty} \gamma^{(2),TT}_na_s^n\,,\notag\\
&  \gamma^{(2),TT}_0= - 4\,,\notag\\
&  \gamma^{(2),TT}_1=-\frac{68}{9}\,,\notag\\
&  \gamma^{(2),TT}_2=-\frac{1412}{81}-\frac{14}{9}\*\zeta_3+\frac{25}{81}\*n_f\,,\notag\\
&  \gamma^{(2),TT}_3=-\frac{2679661}{34992}-\frac{31865}{486}\*\zeta_3+\frac{29}{18}\*\zeta_4+\frac{1955}{27}\*\zeta_5\notag\\
 &\phantom{\gamma^{(2),TT}_3=}+n_f\*\Big(\frac{52337}{17496}+\frac{470}{81}\*\zeta_3-\frac{37}{27}\*\zeta_4\Big)+n_f^2\*\Big(-\frac{95}{8748}+\frac{4}{27}\*\zeta_3\Big)\,.
\end{align}
Additionally, the four loop QCD $\beta$ function is required for the
renormalisation group evolution (Eqs. \eqref{eq:RGE_q}, \eqref{eq:RGE_MStilde}). In our convention
it reads
\begin{align}
  \label{eq:beta}
  &\beta(a_s)=-\sum_{n=0}^{\infty} \beta_na_s^{n+1}\,,\notag\\
&\beta_0=\frac{11}{4}-\frac{n_f}{6}\,,\notag\\
&\beta_1=\frac{51}{8}-\frac{19}{24}n_f\,,\notag\\
&\beta_2=\frac{2857}{128}-\frac{5033}{1152}\*n_f+\frac{325}{3456}\*n_f^2\,,\notag\\
&\beta_3=\frac{149753}{1536}+\frac{891}{64}\*\zeta_3+n_f\*\Big(-\frac{1078361}{41472}-\frac{1627}{1728}\*\zeta_3\Big)\notag\\
 &\phantom{\beta_3=}+n_f^2\*\Big(\frac{50065}{41472}+\frac{809}{2592}\*\zeta_3\Big)+\frac{1093}{186624}\*n_f^3
\,.
\end{align}
The $\beta$ function and the mass anomalous dimension $\gamma_m$ were computed at
four loop order in
Refs. \cite{vanRitbergen:1997va,Czakon:2004bu,Chetyrkin:1997dh,Vermaseren:1997fq}. 

\bibliographystyle{hep_titles}
\bibliography{Q_X_correlators}

\begin{thebibliography}{10}
\providecommand{\url}[1]{\texttt{#1}}
\providecommand{\urlprefix}{URL }
\providecommand{\eprint}[2][]{\url{#2}}

\bibitem{Chetyrkin:1996ia}
K.~G. Chetyrkin, J.~H. {K\"uhn}, A.~Kwiatkowski.
\newblock {QCD corrections to the $e^{+} e^{-}$ cross-section and the $Z$ boson
  decay rate: Concepts and results}.
\newblock \emph{Phys. Rept.}, 277:189--281, 1996.

\bibitem{Chetyrkin:1988yr}
K.~G. Chetyrkin, A.~A. Pivovarov.
\newblock Vacuum saturation hypothesis and qcd sum rules.
\newblock \emph{Nuovo Cim.}, A100:899--906, 1988.
\newblock \eprint{hep-ph/0105093}.

\bibitem{Shuryak:1993kg}
E.~V. Shuryak.
\newblock {Correlation functions in the QCD vacuum}.
\newblock \emph{Rev. Mod. Phys.}, 65:1--46, 1993.

\bibitem{Schafer:1996wv}
T.~Schafer, E.~V. Shuryak.
\newblock {Instantons in QCD}.
\newblock \emph{Rev. Mod. Phys.}, 70:323--426, 1998.
\newblock \eprint{hep-ph/9610451}.

\bibitem{Schafer:2000rv}
T.~Schafer, E.~V. Shuryak.
\newblock {Implications of the ALEPH tau lepton decay data for perturbative and
  non-perturbative QCD}.
\newblock \emph{Phys. Rev. Lett.}, 86:3973--3976, 2001.
\newblock \eprint{hep-ph/0010116}.

\bibitem{Narison:2001ix}
S.~Narison, V.~I. Zakharov.
\newblock {Hints on the power corrections from current correlators in x-space}.
\newblock \emph{Phys. Lett.}, B522:266--272, 2001.
\newblock \eprint{hep-ph/0110141}.

\bibitem{Chu:1993cn}
M.~C. Chu, J.~M. Grandy, S.~Huang, J.~W. Negele.
\newblock {Correlation functions of hadron currents in the QCD vacuum
  calculated in lattice QCD}.
\newblock \emph{Phys. Rev.}, D48:3340--3353, 1993.
\newblock \eprint{hep-lat/9306002}.

\bibitem{DeGrand:2001tm}
T.~A. DeGrand.
\newblock {Short distance current correlators: Comparing lattice simulations to
  the instanton liquid}.
\newblock \emph{Phys. Rev.}, D64:094508, 2001.
\newblock \eprint{hep-lat/0106001}.

\bibitem{Gimenez:2004me}
V.~Gimenez, et~al.
\newblock {Non-perturbative renormalization of lattice operators in coordinate
  space}.
\newblock \emph{Phys. Lett.}, B598:227--236, 2004.
\newblock \eprint{hep-lat/0406019}.

\bibitem{Gimenez:2005nt}
V.~Gimenez, V.~Lubicz, F.~Mescia, V.~Porretti, J.~Reyes.
\newblock Operator product expansion and quark condensate from lattice {QCD} in
  coordinate space.
\newblock \emph{Eur. Phys. J.}, C41:535--544, 2005.
\newblock \eprint{hep-lat/0503001}.

\bibitem{'tHooft:1973mm}
G.~'t~Hooft.
\newblock {Dimensional regularization and the renormalization group}.
\newblock \emph{Nucl. Phys.}, B61:455--468, 1973.

\bibitem{Bardeen:1978yd}
W.~A. Bardeen, A.~J. Buras, D.~W. Duke, T.~Muta.
\newblock {Deep Inelastic Scattering Beyond the Leading Order in Asymptotically
  Free Gauge Theories}.
\newblock \emph{Phys. Rev.}, D18:3998, 1978.

\bibitem{Chetyrkin:1980pr}
K.~G. Chetyrkin, A.~L. Kataev, F.~V. Tkachov.
\newblock {New Approach to Evaluation of Multiloop Feynman Integrals: The
  Gegenbauer Polynomial x Space Technique}.
\newblock \emph{Nucl. Phys.}, B174:345--377, 1980.

\bibitem{Martinelli:1994ty}
G.~Martinelli, C.~Pittori, C.~T. Sachrajda, M.~Testa, A.~Vladikas.
\newblock {A General method for nonperturbative renormalization of lattice
  operators}.
\newblock \emph{Nucl. Phys.}, B445:81--108, 1995.
\newblock \eprint{hep-lat/9411010}.

\bibitem{GorKatLar91:R(s):4l}
S.~G. Gorishny, A.~L. Kataev, S.~A. Larin.
\newblock The {${\cal O} (\alpha_s^3)$} corrections to {$\sigma_{\rm tot} (e^+
  e^- \to {\rm hadrons})$} and {$\Gamma (\tau^- \to \nu_\tau +{\rm hadrons})$}
  in qcd.
\newblock \emph{Phys. Lett.}, B259:144--150, 1991.

\bibitem{SurSam91}
L.~R. Surguladze, M.~A. Samuel.
\newblock {{Total hadronic cross-section in $e^+ e^- $ annihilation at the four
  loop level of perturbative QCD}}.
\newblock \emph{{Phys. Rev. Lett.}}, 66:560--563, 1991.

\bibitem{Chetyrkin:1996sr}
K.~G. Chetyrkin.
\newblock {Correlator of the quark scalar currents and {$\Gamma_{\rm tot}(H\to
  {\rm hadrons})$} at {${\cal O}(\alpha_s^3)$} in pQCD}.
\newblock \emph{Phys. Lett.}, B390:309--317, 1997.
\newblock \eprint{hep-ph/9608318}.

\bibitem{Kiyo:2009gb}
Y.~Kiyo, A.~Maier, P.~{Maierh\"ofer}, P.~Marquard.
\newblock Reconstruction of heavy quark current correlators at {${\cal
  O}(\alpha_s^3)$}.
\newblock \emph{Nucl. Phys.}, B823:269--287, 2009.
\newblock \eprint{arXiv:0907.2120}.

\bibitem{Baikov:2005rw}
P.~A. Baikov, K.~G. Chetyrkin, J.~H. {K\"uhn}.
\newblock {Scalar correlator at ${\cal O}(\alpha_s^4)$, Higgs decay into b-
  quarks and bounds on the light quark masses}.
\newblock \emph{Phys. Rev. Lett.}, 96:012003, 2006.
\newblock \eprint{hep-ph/0511063}.

\bibitem{Baikov:2008jh}
P.~A. Baikov, K.~G. Chetyrkin, J.~H. {K\"uhn}.
\newblock {Order $\alpha^4_s$ {QCD} Corrections to $Z$ and $\tau$ Decays}.
\newblock \emph{Phys. Rev. Lett.}, 101:012002, 2008.
\newblock \eprint{arXiv:0801.1821}.

\bibitem{Gracey:2009da}
J.~A. Gracey.
\newblock {Three loop MSbar operator correlation functions for deep inelastic
  scattering in the chiral limit}.
\newblock \emph{JHEP}, 04:127, 2009.
\newblock \eprint{arXiv:0903.4623}.

\bibitem{Chetyrkin:1997dh}
K.~G. Chetyrkin.
\newblock {Quark mass anomalous dimension to {${\cal O}(\alpha_s^4)$}}.
\newblock \emph{Phys. Lett.}, B404:161--165, 1997.
\newblock \eprint{hep-ph/9703278}.

\bibitem{Vermaseren:1997fq}
J.~A.~M. Vermaseren, S.~A. Larin, T.~van Ritbergen.
\newblock {The 4-loop quark mass anomalous dimension and the invariant quark
  mass}.
\newblock \emph{Phys. Lett.}, B405:327--333, 1997.
\newblock \eprint{hep-ph/9703284}.

\bibitem{Baikov:2006ai}
P.~A. Baikov, K.~G. Chetyrkin.
\newblock {New four loop results in QCD}.
\newblock \emph{Nucl. Phys. Proc. Suppl.}, 160:76--79, 2006.

\bibitem{Chetyrkin:1979bj}
K.~G. Chetyrkin, A.~L. Kataev, F.~V. Tkachov.
\newblock {Higher Order Corrections to {$\sigma_{tot}(e^+ e^- \to
  \mbox{Hadrons})$} in Quantum Chromodynamics}.
\newblock \emph{Phys. Lett.}, B85:277, 1979.

\bibitem{Baikov:prel}
P.~Baikov, K.~Chetyrkin, J.~{K\"uhn}.
\newblock {u}npublished.

\bibitem{Gorishnii:1990zu}
S.~G. Gorishny, A.~L. Kataev, S.~A. Larin, L.~R. Surguladze.
\newblock {CORRECTED THREE LOOP QCD CORRECTION TO THE CORRELATOR OF THE QUARK
  SCALAR CURRENTS AND GAMMA (tot) (H0 $\to$ HADRONS)}.
\newblock \emph{Mod. Phys. Lett.}, A5:2703--2712, 1990.

\bibitem{Baikov:2009uw}
P.~A. Baikov, K.~G. Chetyrkin, J.~H. {K\"uhn}.
\newblock {R(s) and hadronic tau-Decays in Order $\alpha_s^4$: technical
  aspects}.
\newblock 2009.
\newblock \eprint{arXiv:0906.2987}.

\bibitem{Vladimirov:IRR:80}
A.~A. Vladimirov.
\newblock Method for computing renormalization group functions in dimensional
  renormalization scheme.
\newblock \emph{Theor. Math. Phys.}, 43:417, 1980.

\bibitem{Chetyrkin:1984xa}
K.~G. Chetyrkin, V.~A. Smirnov.
\newblock {$R^*$ Operation Corrected}.
\newblock \emph{Phys. Lett.}, B144:419--424, 1984.

\bibitem{vermaseren-form}
J.A.M.Vermaseren.
\newblock New features of \texttt{FORM}.

\bibitem{Gorishnii:1989gt}
S.~G. Gorishny, S.~A. Larin, L.~R. Surguladze, F.~V. Tkachov.
\newblock {MINCER: PROGRAM FOR MULTILOOP CALCULATIONS IN QUANTUM FIELD THEORY
  FOR THE SCHOONSCHIP SYSTEM}.
\newblock \emph{Comput. Phys. Commun.}, 55:381--408, 1989.

\bibitem{Larin:1991fz}
S.~A. Larin, F.~V. Tkachov, J.~A.~M. Vermaseren.
\newblock The \texttt{FORM} version of \texttt{MINCER}.
\newblock NIKHEF-H-91-18.

\bibitem{Chetyrkin:1981qh}
K.~G. Chetyrkin, F.~V. Tkachov.
\newblock Integration by parts: The algorithm to calculate beta functions in 4
  loops.
\newblock \emph{Nucl. Phys.}, B192:159--204, 1981.

\bibitem{Fiamberti:2009jw}
F.~Fiamberti, A.~Santambrogio, C.~Sieg.
\newblock {Five-loop anomalous dimension at critical wrapping order in N=4
  SYM}.
\newblock \emph{JHEP}, 03:103, 2010.
\newblock \eprint{arXiv:0908.0234}.

\bibitem{Baikov:2005nv}
P.~A. Baikov.
\newblock A practical criterion of irreducibility of multi-loop feynman
  integrals.
\newblock \emph{Phys. Lett.}, B634:325--329, 2006.
\newblock \eprint{hep-ph/0507053}.

\bibitem{Baikov:2007zza}
P.~A. Baikov.
\newblock {Recurrence relations in the large space-time dimension limit}.
\newblock \emph{PoS}, RADCOR2007:022, 2007.

\bibitem{Baikov:tadpoles:96}
P.~A. Baikov.
\newblock Explicit solutions of the 3--loop vacuum integral recurrence
  relations.
\newblock \emph{Phys. Lett.}, B385:404--410, 1996.
\newblock \eprint{hep-ph/9603267}.

\bibitem{Baikov:1996cd}
P.~A. Baikov.
\newblock Explicit solutions of the n--loop vacuum integral recurrence
  relations.
\newblock 1996.
\newblock \eprint{hep-ph/9604254}.

\bibitem{Baikov:explit_solutions:97}
P.~A. Baikov.
\newblock Explicit solutions of the multi-loop integral recurrence relations
  and its application.
\newblock \emph{Nucl. Instrum. Meth.}, A389:347--349, 1997.
\newblock \eprint{hep-ph/9611449}.

\bibitem{Baikov:2010hf}
P.~A. Baikov, K.~G. Chetyrkin.
\newblock {Four Loop Massless Propagators: an Algebraic Evaluation of All
  Master Integrals}.
\newblock \emph{Nucl. Phys.}, B837:186--220, 2010.
\newblock \eprint{arXiv:1004.1153}.

\bibitem{Smirnov:2010hd}
A.~V. Smirnov, M.~Tentyukov.
\newblock {Four Loop Massless Propagators: a Numerical Evaluation of All Master
  Integrals}.
\newblock \emph{Nucl. Phys.}, B837:40--49, 2010.
\newblock \eprint{arXiv:1004.1149}.

\bibitem{Fliegner:2000uy}
D.~Fliegner, A.~Retey, J.~A.~M. Vermaseren.
\newblock Parallelizing the symbolic manipulation program form. i: Workstation
  clusters and message passing.
\newblock 2000.
\newblock \eprint{hep-ph/0007221}.

\bibitem{Tentyukov:2004hz}
M.~Tentyukov, et~al.
\newblock {ParFORM: Parallel Version of the Symbolic Manipulation Program
  FORM}.
\newblock 2004.
\newblock \eprint{cs/0407066}.

\bibitem{Tentyukov:2006pr}
M.~Tentyukov, H.~M. Staudenmaier, J.~A.~M. Vermaseren.
\newblock {ParFORM: Recent development}.
\newblock \emph{Nucl. Instrum. Meth.}, A559:224--228, 2006.

\bibitem{Tentyukov:2007mu}
M.~Tentyukov, J.~A.~M. Vermaseren.
\newblock {The multithreaded version of FORM}.
\newblock 2007.
\newblock \eprint{hep-ph/0702279}.

\bibitem{qgraf}
P.~Nogueira.
\newblock Automatic feynman graph generation.
\newblock \emph{J. Comput. Phys.}, 105:279--289, 1993.

\bibitem{vanRitbergen:1997va}
T.~van Ritbergen, J.~A.~M. Vermaseren, S.~A. Larin.
\newblock {The four-loop beta function in quantum chromodynamics}.
\newblock \emph{Phys. Lett.}, B400:379--384, 1997.
\newblock \eprint{hep-ph/9701390}.

\bibitem{Czakon:2004bu}
M.~Czakon.
\newblock The four-loop qcd beta-function and anomalous dimensions.
\newblock \emph{Nucl. Phys.}, B710:485--498, 2005.
\newblock \eprint{hep-ph/0411261}.

\end{thebibliography}


\end{document}